\documentclass[proceedings]{JHEP3}
\usepackage{amsfonts}

\usepackage{epsfig,multicol}

\newbox\mybox

\newcommand\fverb{\setbox\mybox=\hbox\bgroup\verb}
\newcommand\fverbdo{\egroup\medskip\noindent\fbox{\unhbox\mybox}\ }
\newcommand\fverbit{\egroup\item[\fbox{\unhbox\mybox}]}
\conference{Integrable scattering theories with unstable particles}
\abstract{We formulate a new  bootstrap principle which allows for the 
construction of particle spectra involving unstable as well
as stable particles. We comment on the general Lie algebraic 
structure which underlies theories with unstable particles
and propose several new scattering matrices. We
find a new Lie algebraic decoupling rule, which predicts the
renormalization group flow in dependence of the relative ordering
of the resonance parameters. 
The proposals are exemplified for 
some concrete theories which involve unstable particles,
such as  the homogeneous sine-Gordon models
and their generalizations. The new decoupling rule can be validated
by means of  our new bootstrap principle and also via the
renormalization group flow, which we obtain from a 
thermodynamic Bethe ansatz analysis.}

\title{Integrable scattering theories with unstable particles}
\author{O.A.~Castro-Alvaredo, J.~Drei\ss ig and A.~Fring \\
Institut f\"ur Theoretische Physik, Freie Universit\"at Berlin,\\
Arnimallee 14, D-14195 Berlin, Germany \\
E-mail:  \email{olalla/julian.dreissig/fring@physik.fu-berlin.de}}

\input{tcilatex}

\begin{document}

\section{Introduction}

One of the central quantities in the study of quantum field theories is the
scattering matrix $S$, which relates asymptotic in and out states. In
particular in 1+1 space-time dimensions and when concentrating in addition
on integrable theories in this context, the bootstrap principle \cite{boot}
has turned out to be a very powerful non-perturbative construction tool.
Many consistent exact scattering matrices have been determined based on this
idea. Exact is here always to be understood in the sense that $S$ is known
to all orders in perturbation theory. The bootstrap principle was also
successfully generalized to theories in half-space \cite{FK}, theories with
purely transmitting defects \cite{Kon} and even theories which possess
infinitely many resonance states \cite{CF88}. However, hitherto there exists
no formulation of a construction principle leading to unstable particles in
the spectrum. Some specific theories containing unstable particles are
known, but so far the latter emerge as poles in the unphysical sheet as
by-products in the scattering process of two stable particles. A description
of the scattering process of an unstable particle with another stable or
unstable particle is entirely missing in this context. Obviously, scattering
processes involving unstable particles do occur in nature, such that the
quest for a proper prescription is of physical relevance. It is also clear
that this can not be a scattering theory in the usual sense, since for that
one requires the particles involved to exist asymptotically, i.e.~for $%
t\rightarrow \infty $. Clearly any unstable particle will vanish in this
limit rendering such formulation meaningless at first sight. Nonetheless,
some particles have extremely long lifetimes, and appear to exist quasi
infinitely long from an experimentalists point of view. It appears therefore
natural to seek a principle closely related to the conventional bootstrap
for stable particles. One of the main purposes of this paper is to provide
such a construction principle.

Our bootstrap proposal has the following predictive power: a) It yields the
amount of unstable particles together with their mass. This prediction can
be used to explain a mass degeneracy of some unstable particles which can
not be seen in a thermodynamic Bethe ansatz (TBA) analysis. b) It yields the
three-point couplings of all possible interactions, that is, involving
stable as well as unstable particles. c) It is in agreement with a general
Lie algebraic decoupling rule, which we also propose in this paper,
describing the behaviour when certain resonance parameters tend to infinity.

We illustrate the working of the bootstrap for some concrete theories which
are known to contain unstable particles in their spectrum, the homogeneous
sine-Gordon models (HSG) \cite{HSG}. For these models we also test the newly
proposed decoupling rule, which predicts the renormalization group (RG) flow
from the ultraviolet to the infrared directly on the level of the scattering
matrix constructed from the new bootstrap principle and also by means of a
TBA analysis.

\section{A bootstrap for unstable particles}

In general unstable particles of finite life time $\tau $ are described by
complexifying the physical mass of a stable particle, by adding a decay
width $\Gamma \sim \hbar /\tau $. This prescription is well established in
quantum mechanics, see e.g.~\cite{CT} and can also be applied similarly in
the quantum field theory context, see e.g.~\cite{ELOP}. When describing the
latter by means of a scattering theory, the formation of an unstable
particle $(ij)$ from two stable ones of type $i$ and $j$ is well understood.
In that case the creation process is reflected by a pole in the scattering
matrix $S_{ij}$ as a function of the Mandelstam variable at $%
s_{ij}=(m_{(ij)}{}-i\Gamma _{(ij)}/2)^{2}$. As discussed for instance in 
\cite{ELOP}, whenever $m_{(ij)}\gg \Gamma _{(ij)}$, the quantity $m_{(ij)}$
admits a clear cut interpretation as the physical mass. Transforming as
usual in this context from the Mandelstam variable $s$ to the rapidity
variable $\theta $ and describing the scattering of the two stable particles
of type $i$ and $j$ with masses $m_{i}$ and $m_{j}$ by an S-matrix $%
S_{ij}(\theta )$, the resonance pole is situated at $\eta
_{ij}^{(ij)}=\gamma _{ij}^{(ij)}-\QTR{sl}{i}\bar{\gamma}_{ij}^{(ij)}$ with
resonance parameters $\gamma _{ij}^{(ij)},\bar{\gamma}_{ij}^{(ij)}\in \Bbb{R}%
^{+}$. It is crucial to note that the formation of the unstable particle is
accompanied by a parity breaking and poles with $\gamma _{ij}^{(ij)}\in \Bbb{%
R}^{-}$ are not associated to such a process. Identifying the real and
imaginary parts of the pole, then yields the well-known Breit-Wigner
equations \cite{BW} 
\begin{eqnarray}
m_{_{(ij)}}^{2}{}-\frac{\Gamma _{_{ij}}^{2}{}}{4}
&=&m_{i}^{2}{}+m_{j}^{2}{}+2m_{i}m_{j}\cosh \gamma _{ij}^{(ij)}\cos \bar{%
\gamma}_{ij}^{(ij)}  \label{BW1} \\
m_{_{(ij)}}\Gamma _{_{ij}} &=&2m_{i}m_{j}\sinh \gamma _{ij}^{(ij)}\sin \bar{%
\gamma}_{ij}^{(ij)}\,\,.  \label{BW2}
\end{eqnarray}
Eliminating the decay width from (\ref{BW1}) and (\ref{BW2}), we can express
the mass of the unstable particles $m_{_{(ij)}}$ in the model as a function
of the masses of the stable particles $m_{i},m_{j}$ and the resonance
parameter $\gamma _{ij}^{(ij)}$. A simple consequence of (\ref{BW1}) and (%
\ref{BW2}) is that for large resonance parameters $\gamma _{ij}^{(ij)}$ the
masses of the unstable particles are 
\begin{equation}
m_{(ij)}\sim \sqrt{m_{i}m_{j}}e^{\gamma _{ij}^{(ij)}/2}\,,  \label{m}
\end{equation}
which is a relation we will appeal to quite frequently. In some places of
the literature, e.g.~\cite{HSGS,CFKM}, the absolute value of $\gamma
_{ij}^{(ij)}$ is used in (\ref{BW2}) rather than $\gamma _{ij}^{(ij)}$
itself, which seems to suggest that the sign of $\gamma _{ij}^{(ij)}$ is not
relevant. This contradicts, however, the findings of the thermodynamic Bethe
ansatz (TBA) analysis below, in which the signs turn out to be crucial, and
we want to argue here that in fact this absolute value is not needed and
even leads to wrong predictions if implemented. To manifest this point of
view, let us first look at a heuristic argument and consider the
two-particle wave functions $\psi _{ij}\sim \exp (-\Gamma _{_{ij}}t)$ and $%
\psi _{ji}\sim \exp (-\Gamma _{_{ji}}t)$ associated to the scattering of
particles $i$ with $j$ and vice versa. When choosing for definiteness $%
\Gamma _{_{ij}}>0$, the function $\psi _{ij}$ decays for $t\rightarrow
\infty $ and $\psi _{ji}$ decays for $t\rightarrow -\infty $. This is the
effect of parity breaking, suggesting that the unstable particle $(\overline{%
\imath \jmath })$ is formed in the process 
\begin{equation}
i+j\rightarrow (\overline{\imath \jmath })  \label{ij}
\end{equation}
rather than $j+i$. As common also for stable particles we distinguish the
anti-particle of an unstable particle by an overbar, i.e.~$(\overline{\imath
\jmath })$ is the anti-particle of $(ij)$. In general in our notation the
unstable particles can be recognized as they carry the names of their
parents, where the two names are inherited. This interpretation is
compatible with a PT-transformation and hence no absolute value is needed in
(\ref{BW2}). A reason for an artificial introduction of an absolute value is
that apparently in (\ref{BW2}) the parameter $\Gamma $ can become negative
for $\gamma _{ij}^{(ij)}<0$, which is unphysical. However, once one also
attaches indices to the decay width, this concern is eliminated and one
obtains a clear physical meaning for these values. A less heuristic
validation of this point of view will be obtained from our TBA-analysis
presented in section 4.2.

One may now ask the natural question whether one can cross the unstable
particle to the other side in the process (\ref{ij}), that is, do the
processes $(ij)+i$ or $\ j+(ij)$ make any sense and, furthermore, is it
possible to formulate a bootstrap principle for unstable particles? To
formulate such a principle is highly desirable, since it would allow for an
explicit construction of unstable particles. Up to now they only emerge
indirectly, somewhat as a side product once the stable particle content has
been determined.

As already mentioned in the introduction, the main conceptual obstacle in
the formulation of a bootstrap principle for unstable particles is the fact
that in a well defined scattering theory one always deals with asymptotic
states. Nonetheless, one may seek an approach closely related to a properly
defined S-matrix.

Let us formulate the bootstrap principle: We commence with the fusing of two
stable particles to create an unstable particle as in the process (\ref{ij}%
). To the process (\ref{ij}) we can associate bootstrap equations almost in
the usual way. We scatter for this with an additional particle, say of type $%
l$, and exploit the integrability of the theory. Accordingly the ordering of
the scattering is associative, such that we can equate the two situations of
either $l$ scattering before or after the creation of the unstable particle.
The object which describes the latter process we refer to as $\tilde{S}%
_{l(ij)}$, indicating with the tilde that we do not view this object as a
standard S-matrix since it involves one particle which is unstable. In the
conventional formulation one generally assumes that also the created
particle exists asymptotically. It is this assumption we propose to relax.
We depict the bootstrap equation

\FIGURE{\epsfig{file=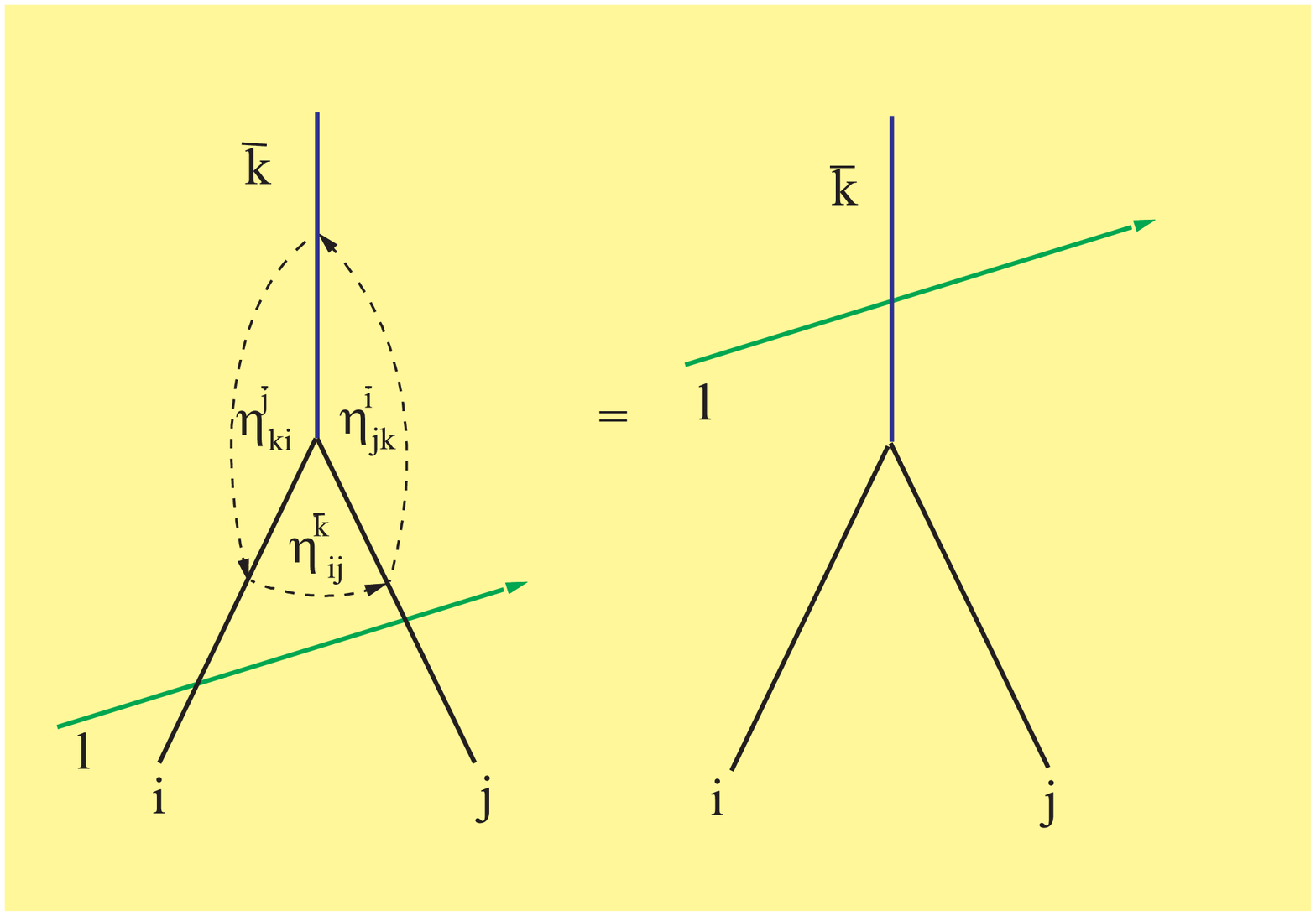,width=7.5cm,height=5.18cm} 
        \caption{The Bootstrap equations.}        \label{figure1}}

\noindent in figure 1. Identifying in figure 1 the particle $k$ with $(ij)$
we obtain, according to the outlined conventions, the $\tilde{S}$ bootstrap
equations 
\begin{equation}
\tilde{S}_{li}(\theta -\bar{\eta}_{(ij)i}^{\overline{\jmath }})\,\tilde{S}%
_{lj}(\theta +\bar{\eta}_{j(ij)}^{\overline{\imath }})\,=\tilde{S}_{l(%
\overline{\imath \jmath })}(\theta ),  \label{b2}
\end{equation}
where $\bar{\eta}=\pm \QTR{sl}{i}\pi -$ $\eta $ and also $\bar{\eta}%
\rightarrow -\bar{\eta}$ is not a symmetry. In figure 1 we indicate that the
angles should be measured anti-clockwise, which explains the signs. We also
note that we do not assume parity invariance, such that in general $\bar{\eta%
}_{ji}^{(\overline{\imath \jmath })}\neq \bar{\eta}_{ij}^{(\overline{\imath
\jmath })}$. As further analogy between $\tilde{S}$ and $S$, we assume that
the crossing and analyticity relations are maintained 
\begin{equation}
\tilde{S}_{kl}(\theta )=\tilde{S}_{\bar{l}k}(\QTR{sl}{i}\pi -\theta )\qquad 
\text{and}\qquad \tilde{S}_{kl}(\theta )\tilde{S}_{lk}(-\theta )=1\,.
\label{c}
\end{equation}
However, we do not demand $\tilde{S}$ to be unitary, for reasons we will
comment upon further below. With the help of (\ref{c}), one derives the
bootstrap equations for the opposite parity and the ones for the crossed
processes $i+(ij)\rightarrow \bar{\jmath}$ and $j+(ij)\rightarrow \bar{\imath%
}$ \ from (\ref{b2}) 
\begin{eqnarray}
\tilde{S}_{(\overline{\imath \jmath })l}(\theta ) &=&\tilde{S}_{il}(\theta +%
\bar{\eta}_{(ij)i}^{\overline{\jmath }})\,\tilde{S}_{jl}(\theta -\bar{\eta}%
_{j(ij)}^{\overline{\imath }})\,,  \label{b33} \\
\tilde{S}_{l\bar{\jmath}}(\theta ) &=&\,\tilde{S}_{l(ij)}(\theta -\bar{\eta}%
_{j(ij)}^{\overline{\imath }})\,\tilde{S}_{li}(\theta \pm \QTR{sl}{i}\pi -%
\bar{\eta}_{(ij)i}^{\bar{\jmath}}-\bar{\eta}_{j(ij)}^{\overline{\imath }}),
\label{b3} \\
\tilde{S}_{l\bar{\imath}}(\theta ) &=&\tilde{S}_{l(ij)}(\theta +\bar{\eta}%
_{(ij)i}^{\bar{\jmath}})\tilde{S}_{lj}(\theta \pm \QTR{sl}{i}\pi +\bar{\eta}%
_{(ij)i}^{\bar{\jmath}}+\bar{\eta}_{j(ij)}^{\overline{\imath }})\,\,.
\label{b4}
\end{eqnarray}
From the crossing relation for the scattering matrix and (\ref{b3}) or (\ref
{b4}) one obtains some relations between the various fusing angles 
\begin{equation}
\bar{\eta}_{ij}^{(\overline{\imath \jmath })}+\bar{\eta}_{(ij)i}^{\bar{\jmath%
}}+\bar{\eta}_{j(ij)}^{\overline{\imath }}=\pm \QTR{sl}{i}\pi \,\,.
\label{eta}
\end{equation}
At first sight this looks very much like the usual bootstrap prescription,
but there are some differences. As is clear from the scattering process of
two stable particles producing an unstable one, the angle $\bar{\eta}_{ij}^{(%
\overline{\imath \jmath })}$ is not purely complex any longer as it is for
the situation when exclusively stable particles scatter. As a consequence,
this property then extends to the other angles $\bar{\eta}_{(ij)i}^{\bar{%
\jmath}}$ and $\bar{\eta}_{j(ij)}^{\overline{\imath }}$ in (\ref{b2}), which
also possess some non-vanishing real parts. Note that (\ref{eta}) implies
that the real parts of the three angels involved add up to zero. At this
point we do not have an entirely compelling reason for demanding that, but
this formulation will turn out to work well. The main conceptual difference
is of course that we allow unstable particles to be involved in the
scattering processes. In the time interval $0<t<\tau _{(ij)}$ we could
formally associate to those particles some operators $\tilde{Z}%
_{(ij)}^{\dagger }(\theta )$, with $\lim_{t\rightarrow \infty }\tilde{Z}%
_{(ij)}^{\dagger }(\theta )=1$ if $\tau _{(ij)}<\infty $. It is important to
note that these operators do not exist asymptotically, such that in general $%
\left\langle \tilde{Z}_{(ij)}(\theta )\tilde{Z}_{(ij)}^{\dagger }(\theta
^{\prime })\right\rangle $ $\neq \delta (\theta -\theta ^{\prime })$. In
fact the entire operation of crossing $\tilde{Z}_{(ij)}^{\dagger }(\theta )$
from an in to an out state is ill defined. This property is then reflected
in the non-unitary of $\tilde{S}$. The loss of unitarity is somewhat
natural, since it usually reflects conservation of probabilities, which of
course we do not expect in the case of unstable particles.

It can already be anticipated that our proposal will lead to the
construction of further unstable particles besides the primary (fundamental)
ones created in the scattering process of two stable ones. In the following
we shall refer to unstable particles formed in a scattering process
involving at least one primary unstable particle as ``secondaries'' and to
those which have at least one secondary as their parent as ``tertiaries'',
etc. An important observation will be that tertiaries can be degenerate in
mass to stables, primaries or secondaries.

\section{Generalities on theories with unstable particles}

At present all known theories with unstable particles in their spectrum can
be thought of in terms of a simple universal Lie algebraic structure. The
formulation is based on an arbitrary simply laced Lie algebra \textbf{\~{g}}
(possibly with a subalgebra \textbf{\~{h}}) with rank $\tilde{\ell}$
together with its associated Dynkin diagram (for more details see for
instance \cite{Hum}). To each node one attaches a simply laced Lie algebra 
\textbf{g}$_{i}$ and to each link between the nodes $i$ and $j$ a resonance
parameter $\sigma _{ij}$, as depicted in the following \textbf{\~{g}/\~{h}}%
-coset Dynkin diagram

\unitlength=1.0cm 
\begin{picture}(14.20,1.4)(-3.8,-0.7)
\put(-1.,0.00){\circle*{0.2}}

\put(-1.25,-0.50){${\bold g}_1$}

\put(-0.7,0.00){$\ldots$}
\put(0,0.00){\circle*{0.2}}
\put(0.27,0.30){$ \sigma_{ij}$}

\put(-0.25,-0.50){${\bold g}_i$}
\put(0.00,-0.01){\line(1,0){0.9}}
\put(0.35,-0.01){\line(2,1){0.4}}
\put(0.35,-0.01){\line(2,-1){0.4}}
\put(1.00,0.00){\circle*{0.2}}

\put(1.35,0.30){$ \sigma_{jk}$}
\put(0.9,-0.50){${\bold g}_j$}

\put(1.0,-0.01){\line(1,0){0.9}}
\put(1.65,-0.01){\line(-2,1){0.4}}
\put(1.65,-0.01){\line(-2,-1){0.4}}
\put(2.0,0.00){\circle*{0.2}}

\put(1.9,-0.50){${\bold g}_k$}
\put(2.2,0.00){$\ldots$}

\put(3.0,0.00){\circle*{0.2}}

\put(2.9,-0.50){${\bold g}_{\tilde \ell }$}

\put(3.5,-.6){\line(1,1){1.0}}

\put(4.5,0.00){$\ldots$}
\put(5.2,0.00){\circle*{0.2}}

\put(5.55,0.30){$ \sigma_{lm}$}

\put(4.95,-0.50){${\bold g}_l$}
\put(5.20,-0.01){\line(1,0){0.9}}
\put(5.55,-0.01){\line(2,1){0.4}}
\put(5.55,-0.01){\line(2,-1){0.4}}
\put(6.20,0.00){\circle*{0.2}}
\put(6.45,0.30){$ \sigma_{mn}$}

\put(6.1,-0.50){${\bold g}_m$}

\put(6.2,-0.01){\line(1,0){0.9}}
\put(6.85,-0.01){\line(-2,1){0.4}}
\put(6.85,-0.01){\line(-2,-1){0.4}}
\put(7.2,0.00){\circle*{0.2}}
\put(7.1,-0.50){${\bold g}_n$}
\put(7.4,0.00){$\ldots$}

\end{picture}

\noindent Besides the usual rules for Dynkin diagrams, we adopt here the
convention that we add an arrow to the link, which manifests the parity
breaking and allows to identify the signs of the resonance parameters. An
arrow pointing from the node $i$ to $j$ simply indicates that $\sigma
_{ij}>0 $. Since, except in section 4.2.6, we are dealing mostly with simply
laced Lie algebras, this should not lead to confusion.

As particular examples one can choose for instance \textbf{\~{g} }to be
simply laced and $\mathbf{g}_{1}=\ldots = \mathbf{g}_{\tilde{\ell}}=SU(k)$,
in which case one obtains the \textbf{\~{g}}$_{k}$-homogeneous sine-Gordon
models \cite{HSG,HSGS}. This is generalized \cite{CK} when taking instead 
\textbf{\~{g} }to be non-simply laced and \textbf{g}$_{i}=SU(2k/\alpha
_{i}^{2})$, with $\alpha _{i}$ being the simple roots of \textbf{\~{g}}. The
choice $\mathbf{g}_{1}=\ldots =\mathbf{g}_{\tilde{\ell}}=\mathbf{g }$ with 
\textbf{g} being any arbitrary simply laced Lie algebra gives the \textbf{g%
\TEXTsymbol{\vert}\~{g}}-theories \cite{AFCK}. An example for a theory
associated to a coset is the roaming sinh-Gordon model \cite{Stair}, which
can be thought of as \textbf{\~{g}/\~{h}}$\equiv \lim_{k\rightarrow \infty
}SU(k+1)/SU(k)$ with $\mathbf{g}_{1}=\ldots =\mathbf{g}_{\tilde{\ell} }=SU(2)
$. It is clear that the examples presented here do not exhaust yet all
possible combinations and the structure mentioned above allows for more
combinations of algebras, which are not yet explored. One is also not
limited to Dynkin diagrams and may consider more general graphs which have
multiple links, i.e.~resonance parameters, between various nodes. Examples
for such theories were proposed and studied in \cite{CF88}.

On the base of this Lie algebraic picture one can then easily construct the
scattering matrix. For this, we characterize each particle by two quantum
numbers $(a,i)$, which take their values in different ranges $1\leq a\leq
\ell _{i}$, where $\ell _{i}$ is not necessarily rank \textbf{g}$_{i}$ and $%
1\leq i\leq \tilde{\ell}=$rank \textbf{\~{g}}. This means, in total we have $%
(\tilde{\ell}\times \sum_{i}\ell _{i})$ different particle types. The
scattering matrix describing the interaction between these type of particles
is then of the general form 
\begin{equation}
S_{ab}^{ij}(\theta ,\sigma _{ij})=[S_{ab}^{\min }(\theta )]^{\delta
_{ij}}[S_{ab}^{ij}(\theta ,\sigma _{ij})]^{\tilde{I}_{ij}}\,,
\end{equation}
where $S_{ab}^{\min }(\theta )$ corresponds to some scaling model of a
statistical model and $\tilde{I}$ is the incidence matrix of \textbf{\~{g}}.
It is known that all S-matrices of these scaling models can be viewed as
minimal parts of some affine Toda field theory (ATFT) S-matrix \cite{TodaS}.
Then $S_{ab}^{ij}(\theta ,\sigma _{ij})$ is essentially a CDD-factor \cite
{CDD} for which the coupling constant is chosen to be a function of the
values $i,j$. At present, all known scattering theories which involve
unstable particles are of this generic form and in particular all of the
above mentioned examples.

We propose here yet a further class of scattering matrices. The presented
picture makes it very suggestive to be generalized by introducing a coupling
constant dependence into such models simply by replacing $S_{ab}^{\min
}(\theta )$ by a full affine Toda field theory scattering matrix 
\begin{equation}
S_{ab}^{ij}(\theta ,B,\sigma _{ij}):=[S_{ab}^{ATFT}(\theta ,B)]^{\delta
_{ij}}[S_{ab}^{ij}(\theta ,\sigma _{ij})]^{\tilde{I}_{ij}}\,.  \label{BHSG}
\end{equation}
Obviously this will not alter any of the bootstrap consistency equations,
since the difference between the theory with a minimal S-matrix and the one
with a coupling constant dependence such as (\ref{BHSG}) is simply a CDD
factor. It should also be clear that we may introduce an additional coupling
constant into the second factor in (\ref{BHSG}) by a similar consideration.
As an illustration of the construction principle (\ref{BHSG}) we present
here a new scattering matrix which corresponds to the coupling constant
dependent version of the $SU(3)_{2}$-HSG model. The two self-conjugate
particles in the model named by $``+"$ and $``-"$ interact as 
\begin{equation}
S_{\pm \pm }(\theta ,B)=\frac{\tanh \frac{1}{2}\left( \theta -i\frac{\pi }{2}%
B\right) }{\tanh \frac{1}{2}\left( \theta +i\frac{\pi }{2}B\right) }\quad
\quad \text{and}\quad \quad S_{\pm \mp }(\theta ,\sigma )=\pm \tanh \frac{1}{%
2}\left( \theta \pm \sigma -i\frac{\pi }{2}\right) \,\,.\;  \label{S}
\end{equation}
It is easy to check that this matrices satisfy the unitarity-analyticity and
crossing relations. There is no pole in the physical sheet, such that no
fusing takes place in this model. There exist various distinct limits:
Taking the coupling constant $B$ to be real and $0\leq B\leq 2$ the
scattering of two particles of the same type is described by the sinh-Gordon
scattering matrix. The analytic continuation $B\rightarrow 1+i\bar{\sigma}%
/\pi $ turns $S_{\pm \pm }(\theta )$ into two copies of the roaming
trajectory S-matrix of Zamolodchikov \cite{Stair}. Taking the limit $\bar{%
\sigma}\rightarrow \infty $ reduces the whole system (\ref{S}) to the usual
SU(3)$_{2}$-HSG model. The limit $\sigma \rightarrow \infty $ decouples the
system into a direct product of two theories described either by the
sinh-Gordon scattering matrix, two roaming trajectory S-matrices, two
thermally perturbed Ising models or a free boson, depending on the value of $%
B$.

\subsection{Decoupling rule}

Of special interest is to investigate the behaviour of such systems when
certain resonance parameters $\sigma _{ij}$ tend to infinity. According to (%
\ref{m}) this limit corresponds to a situation when the energy scale of the
unstable particle is so large that it can never be created. As a
consequence, the particle content of the theory will be altered when such a
limit is carried out. This feature should be captured by the bootstrap
construction. When taking several of such limits in a consecutive order,
this behaviour is reflected in addition in the renormalization group flow in
form of the typical staircase pattern of the Virasoro central charge as a
function of the inverse temperature as will be discussed below. In the
direction from the infrared to the ultraviolet, the flow from one plateau to
the next is then associated to the formation of an unstable particle with
mass (\ref{m}). The challenge is of course to predict the positions, that
is, the height and the on-set of the plateaux, as a function of the
temperature. The on-set is simply determined by the formula (\ref{m}). In
order to predict the height, i.e.~the fixed points of the RG flow, we
propose the following

\smallskip \indent {\rm{Decoupling rule}:} \emph{Call the overall Dynkin
diagram }$\mathcal{C}$\emph{\ and denote the associated Lie group and Lie
algebra by }$\tilde{G}_{\mathcal{C}}$\emph{\ and \textbf{\~{g}}}$_{\mathcal{C%
}}$\emph{, respectively. Let }$\sigma _{ij}$\emph{\ be some resonance
parameter related to the link between the nodes }$i$\emph{\ and }$j$\emph{.
To each node }$i$\emph{\ attach a simply laced Lie algebra }$\mathbf{g}_{i}.$%
\emph{\ Produce a reduced diagram }$\mathcal{C}_{ji}$\emph{\ containing the
node }$j$\emph{\ by cutting the link adjacent to it in the direction }$i$%
\emph{. Likewise produce a reduced diagram }$\mathcal{C}_{ij}$\emph{\
containing the node }$i$\emph{\ by cutting the link adjacent to it in the
direction }$j$\emph{. Then the }$\tilde{G}_{\mathcal{C}}$\emph{-theory
decouples according to the rule} 
\begin{equation}
\lim_{\sigma _{ij}\rightarrow \infty }\tilde{G}_{\mathcal{C}}=\tilde{G}_{(%
\mathcal{C}-\mathcal{C}_{ij})}\otimes \tilde{G}_{(\mathcal{C}-\mathcal{C}%
_{ji})}/\tilde{G}_{(\mathcal{C}-\mathcal{C}_{ij}-\mathcal{C}_{ji})}\,.
\label{drule}
\end{equation}
We depict this rule also graphically in terms of Dynkin diagrams:

\unitlength=1.0cm 
\begin{picture}(15.20,3.0)(0,-2.00)
\put(-0.7,0.00){$\ldots$}
\put(0.00,0.00){\circle*{0.2}}
\put(-0.10,-0.50){$ {\bold g}_i$}
\put(0.00,-0.01){\line(1,0){0.9}}
\put(1.00,0.00){\circle{0.2}}
\put(1.80,0.30){${\cal  C}$}
\put(1.10,-0.01){\line(1,0){0.5}}
\put(1.6,0.00){$\ldots$}
\put(2.2,-0.01){\line(1,0){0.4}}
\put(2.70,0.00){\circle{0.2}}
\put(2.80,-0.01){\line(1,0){1.0}}
\put(3.70,0.00){\circle*{0.2}}
\put(3.60,-0.50){${\bold g}_{j}$}
\put(3.90,-0.01){$\ldots$ }
\put(4.70,-0.1){$\Rightarrow$}

\put(5.3,0.00){$\ldots$}
\put(6.00,0.00){\circle*{0.2}}
\put(5.9,-0.50){${\bold g}_i$}
\put(6.00,-0.01){\line(1,0){0.9}}
\put(7.00,0.00){\circle{0.2}}
\put(6.9,0.30){${\cal  C}-{\cal C}_{ji}$}
\put(7.10,-0.01){\line(1,0){0.5}}
\put(7.6,0.00){$\ldots$}
\put(8.2,-0.01){\line(1,0){0.4}}
\put(8.70,0.00){\circle{0.2}}

\put(4.,0.30){ \small $ \sigma_{ij} \rightarrow \infty  $}

\put(9.20,-0.1){$\otimes$}

\put(10.00,0.00){\circle{0.2}}
\put(10.80,0.30){${\cal  C} -{\cal C}_{ij }$}
\put(10.10,-0.01){\line(1,0){0.5}}
\put(10.6,0.00){$\ldots$}
\put(11.2,-0.01){\line(1,0){0.4}}
\put(11.70,0.00){\circle{0.2}}
\put(11.80,-0.01){\line(1,0){1.0}}
\put(12.70,0.00){\circle*{0.2}}
\put(12.60,-0.50){${\bold g}_{j}$}
\put(12.90,-0.01){$\ldots$ }

\put(13.4,-.6){\line(1,1){1.0}}

\put(12.00,-2.00){\circle{0.2}}
\put(11.90,-1.70){${\cal  C} -{\cal C}_{ij} -{\cal C}_{ji} $}
\put(12.10,-1.99){\line(1,0){0.5}}
\put(12.6,-2.00){$\ldots$}
\put(13.2,-1.99){\line(1,0){0.4}}
\put(13.70,-2.00){\circle{0.2}}

\end{picture}

\noindent According to the GKO-coset construction \cite{GKO}, this means
that the Virasoro central charge flows as 
\begin{equation}
c_{\mathbf{\tilde{g}}_{\mathcal{C}}}\rightarrow c_{\mathbf{\tilde{g}}_{%
\mathcal{C-C}_{ij}}}+c_{\mathbf{\tilde{g}}_{\mathcal{C-C}_{ji}}}-c_{\mathbf{%
\tilde{g}}_{\mathcal{C}-\mathcal{C}_{ij}-\mathcal{C}_{ji}}}\,\,.  \label{dc}
\end{equation}
The rule may be applied consecutively to each disconnected subgraph produced
according to the decoupling rule. We stress that this rule is really a
decoupling rule and not a fusing rule, it only predicts the flow from the
ultraviolet to the infrared and not vice versa. This is of course natural as
scaling functions measure the degrees of freedom of the system and the
information loss in the RG flow is irreversible. Hence, potentially we have
for the visible unstable particles $[\ell (\ell -1)/2]!$ different types of
flows corresponding to the possible orderings for the masses. For $SU(\ell
+1)$ the particular ordering $\sigma _{2i,2i+1}>0$, $\sigma _{2i-1,2i}<0$
with $|\sigma _{2i,2i+1}|<|\sigma _{2i-1,2i}|$ leads to the decoupling rule
as discussed in \cite{CF8}, which is a special case of (\ref{drule}). In the
following we shall confirm the general rule (\ref{drule}) directly on the
level of the scattering matrix and also by means of a TBA analysis which
yields the RG flow passing along the central charges as predicted by (\ref
{dc}).

More familiar is a decoupling rule found by Dynkin \cite{Dynkin} for the
construction of semi-simple\footnote{%
The subalgebras constructed in this way are not necessarily maximal and
regular. A guarantee for obtaining those, exept in six special cases, is
only given when one manipulates adequately the extended Dynkin diagram.}
subalgebras \textbf{\~{h}} from a given algebra \textbf{\~{g}}. For the more
general diagrams which can be related to the \textbf{\~{g}}$_{k}$-HSG models
the generalized rule can be found in \cite{Kuniba}. These rules are all
based on removing some of the nodes rather than links. For our physical
situation at hand this corresponds to sending all stable particles which are
associated to the algebra of a particular node to infinity. As in the
decoupling rule (\ref{drule}) the number of stable particles remains
preserved, it is evident that the two rules are inequivalent. Letting for
instance the mass scale in \textbf{g}$_{j}$ go to infinity, the generalized
(in the sense that \textbf{g}$_{j}$ can be different from $A_{\ell }$) rule
of Kuniba is simply depicted as

\unitlength=1.0cm 
\begin{picture}(15.20,2.0)(-2.0,-1.00)
\put(-0.7,0.00){$\ldots$}
\put(0.00,0.00){\circle*{0.2}}
\put(-0.10,-0.50){$ {\bold g}_i$}
\put(0.00,-0.01){\line(1,0){0.9}}
\put(1.00,0.00){\circle*{0.2}}

\put(0.90,-.50){$ {\bold g}_j$}

\put(0.9,0.30){${\cal  C}$}
\put(1.10,-0.01){\line(1,0){0.9}}
\put(2.00,0.00){\circle*{0.2}}
\put(1.90,-0.50){$ {\bold g}_k$}
\put(2.20,-0.01){$\ldots$ }

\put(3.80,-0.1){$\Rightarrow$}
\put(3.1,0.30){ \small $ m_j \rightarrow \infty  $}

\put(5.3,0.00){$\ldots$}
\put(6.00,0.00){\circle*{0.2}}
\put(5.9,-0.50){${\bold g}_i$}
\put(5.5,0.30){${\cal  C}-{\cal C}_{ji}$}

\put(7.50,-0.1){$\otimes$}

\put(8.50,0.30){${\cal  C} -{\cal C}_{jk} $}

\put(9.0,0.00){\circle*{0.2}}
\put(8.90,-0.50){${\bold g}_{k}$}
\put(9.20,-0.01){$\ldots$ }
\end{picture}

We illustrate the working of the rule (\ref{drule}) with the simple example
of the $SU(4)_{2}$-HSG model. For level 2, we can associate the simple roots
to the nodes of the \textbf{\~{g}}-Dynkin diagram. For the ordering $\sigma
_{13}>\sigma _{12}>\sigma _{23}$ we predict from (\ref{drule}) the flow

\unitlength=1.0cm 
\begin{picture}(15.20,4.0)(-1.,-2.5)
\put(0,1.00){\circle*{0.2}}
\put(-0.25,.50){$\alpha_1$}
\put(0.00,0.99){\line(1,0){0.9}}
\put(1.00,1.00){\circle*{0.2}}
\put(0.9,0.50){$\alpha_2$}
\put(1.00,0.99){\line(1,0){0.9}}
\put(2.00,1.00){\circle*{0.2}}
\put(1.9,0.50){$\alpha_3$}
\put(7.3,0.85){\footnotesize $G = SU(4)_2$}
\put(11.3,0.8){$c = 2$}
\put(-1.5,0.00){$ \rightarrow \sigma_{13}$}
\put(0,0.00){\circle*{0.2}}
\put(-0.25,-0.50){$\alpha_1$}
\put(0.00,-0.01){\line(1,0){0.9}}
\put(1.00,0.00){\circle*{0.2}}
\put(0.9,-0.50){$\alpha_2$}
\put(1.50,-0.1){$\otimes$}
\put(2.1,0.0){\circle*{0.2}}
\put(1.95,-0.50){$\alpha_2$}
\put(2.10,-0.01){\line(1,0){0.9}}
\put(3.10,0.00){\circle*{0.2}}
\put(3.,-0.50){$\alpha_3$}
\put(3.6,-.6){\line(1,1){1.0}}
\put(4.90,0.00){\circle*{0.2}}
\put(4.8,-.50){$\alpha_2$}
\put(7.3,-0.15){\footnotesize $G = SU(3)^{\otimes 2}_2/SU(2)_2$}
\put(11.3,-0.2){$c = 1.9$}
\put(-1.5,-1.00){$ \rightarrow  \sigma_{12}$}
\put(.00,-1.00){\circle*{0.2}}
\put(-0.25,-1.50){$\alpha_1$}
\put(.50,-1.1){$\otimes$}
\put(1.1,-1.0){\circle*{0.2}}
\put(0.95,-1.50){$\alpha_2$}
\put(1.10,-1.01){\line(1,0){0.9}}
\put(2.10,-1.00){\circle*{0.2}}
\put(2.,-1.50){$\alpha_3$}
\put(7.3,-1.15){\footnotesize $G = SU(3)_2 \otimes SU(2)_2 $} 
\put(11.3,-1.2){$c = 1.7$}
\put(-1.5,-2.00){$ \rightarrow \sigma_{23}$}
\put(0,-2.00){\circle*{0.2}}
\put(-0.25,-2.50){$\alpha_1$}
\put(.40,-2.1){$\otimes$}
\put(1.00,-2.00){\circle*{0.2}}
\put(0.9,-2.50){$\alpha_2$}
\put(1.50,-2.1){$\otimes$}
\put(2.1,-2.0){\circle*{0.2}}
\put(1.95,-2.50){$\alpha_3$}
\put(7.3,-2.15){\footnotesize $G = SU(2)^{\otimes 3}_2 $}
\put(11.3,-2.2){$c = 1.5$}
\end{picture}

\noindent Taking instead the ordering $\sigma _{23}>\sigma _{13}>\sigma
_{12} $, we compute

\begin{picture}(15.20,3.8)(-1.,-2.5)

\put(0,1.00){\circle*{0.2}}
\put(-0.25,.50){$\alpha_1$}
\put(0.00,0.99){\line(1,0){0.9}}
\put(1.00,1.00){\circle*{0.2}}
\put(0.9,0.50){$\alpha_2$}
\put(1.00,0.99){\line(1,0){0.9}}
\put(2.00,1.00){\circle*{0.2}}
\put(1.9,0.50){$\alpha_3$}
\put(7.3,0.85){\footnotesize $G = SU(4)_2$}
\put(11.3,0.8){$c = 2$}
\put(-1.5,0.00){$\rightarrow \sigma_{23}$}
\put(0,0.00){\circle*{0.2}}
\put(-0.25,-0.50){$\alpha_1$}
\put(0.00,-0.01){\line(1,0){0.9}}
\put(1.00,0.00){\circle*{0.2}}
\put(0.9,-0.50){$\alpha_2$}
\put(1.50,-0.1){$\otimes$}
\put(2.1,0.0){\circle*{0.2}}
\put(1.95,-0.50){$\alpha_3$}
\put(7.3,-0.15){\footnotesize $G = SU(3)_2 \otimes SU(2)_2$}
\put(11.3,-0.2){$c = 1.7$}
\put(-1.5,0.00){$\rightarrow \sigma_{23}$}
\put(0,0.00){\circle*{0.2}}
\put(-0.25,-0.50){$\alpha_1$}
\put(0.00,-0.01){\line(1,0){0.9}}
\put(1.00,0.00){\circle*{0.2}}
\put(0.9,-0.50){$\alpha_2$}
\put(1.50,-0.1){$\otimes$}
\put(2.1,0.0){\circle*{0.2}}
\put(1.95,-0.50){$\alpha_3$}
\put(7.3,-0.15){\footnotesize $G = SU(3)_2 \otimes SU(2)_2$}
\put(11.3,-0.2){$c = 1.7$}
\put(-1.5,-1.00){$ \rightarrow \sigma_{13}$}
\put(.00,-1.025){is already decoupled}
\put(-1.5,-2.00){$ \rightarrow \sigma_{12}$}
\put(0,-2.00){\circle*{0.2}}
\put(-0.25,-2.50){$\alpha_1$}
\put(.40,-2.1){$\otimes$}
\put(1.00,-2.00){\circle*{0.2}}
\put(0.9,-2.50){$\alpha_2$}
\put(1.50,-2.1){$\otimes$}
\put(2.1,-2.0){\circle*{0.2}}
\put(1.95,-2.50){$\alpha_3$}
\put(7.3,-2.15){\footnotesize $G = SU(2)^{\otimes 3}_2 $}
\put(11.3,-2.2){$c = 1.5$}

\end{picture}

\noindent The central charges can be obtained from (\ref{cdel}) using (\ref
{dc}). We will elaborate more on this example in section 4.1.2 and 4.2.1
from the bootstrap and TBA point of view, respectively. It is important to
note the non-commutative nature of the limiting procedures. For more
complicated algebras it is essential to keep track of the labels on the
nodes, since only in this way one can decide whether they cancel against the
subgroup diagrams or not. See section 4.2 for more details and examples.

\section{The homogeneous sine-Gordon models}

Let us now consider some specific choice of algebras being\ $\mathbf{g}%
_{1}=\ldots =\mathbf{g}_{\tilde{\ell}}=SU(k)$, which as we mentioned
corresponds to the \textbf{\~{g}}$_{k}$-HSG models. These theories have
recently attracted some attention, because it could be argued even from a
Lagrangian point of view that besides stable particles they also contain
unstable particles in their spectrum \cite{HSG}. The HSG-models belong to
the huge class of massive integrable models which can be obtained as
relevant perturbations from conformal field theories in the spirit of \cite
{Zamo}, 
\begin{equation}
\mathcal{H}_{G_{k}\text{-HSG}}=\mathcal{H}_{G_{k}/U(1)^{\ell }\text{-CFT}%
}-\lambda \int d^{2}x\phi (x,t)\,.  \label{pert}
\end{equation}
The underlying conformal field theory is a WZNW-$G_{k}/U(1)^{\ell }$-coset
theory \cite{Witten,GKO} with $k$ being the level and $\ell $ the rank of a
semi-simple Lie algebra \textbf{g}. The Virasoro central charge $c\,\ $and
the conformal dimensions $\Delta ,\bar{\Delta}$ of the perturbing operator $%
\phi $ are 
\begin{equation}
c=\ell \,\frac{k\,h-h^{\vee }}{k+h^{\vee }}\qquad \text{and\qquad }\Delta =%
\bar{\Delta}=\frac{h^{\vee }}{k+h^{\vee }}\,.  \label{cdel}
\end{equation}
Here $(h^{\vee })\,h$ is the (dual) Coxeter number of \textbf{g}. In the
notation of the massive theory we will not carry along the subalgebra $%
U(1)^{\ell }$ as indicated in (\ref{pert}) for simplicity. The S-matrices
involving the stable particles for simply laced and non-simply laced
algebras were proposed in \cite{HSGS} and \cite{CK}, respectively. Examples
on which we will focus a lot in the following are models with level $k=2$.
Adopting the notation of \cite{CF8}, the related scattering matrix may be
written for the simply laced case as 
\begin{equation}
S_{ij}(\theta ,\sigma _{ij})=(-1)^{\delta _{ij}}\,\varepsilon (\sigma
_{ij})(\sigma _{ij},2)^{I_{ij}}\,,\qquad 1\leq i,j\leq \ell   \label{SHSG}
\end{equation}
where $I$ denotes the incidence matrix of \textbf{g. }It is convenient to
abbreviate 
\begin{equation}
(\sigma ,x):=\tanh (\theta +\sigma -i\pi x/4)/2\,.  \label{building}
\end{equation}

\noindent The $\varepsilon (x)$ is the step-function, i.e.~$\varepsilon
(x)=1 $ for $x\geq 0$, $\varepsilon (x)=-1$ for $x<0$. The model contains $%
(\ell -1)$ linear independent resonance parameters $\sigma _{ij}$. As a
convenient basis one usually chooses those which can be associated directly
to the links in the Dynkin diagram, i.e.~the primary unstable particles. The 
$\sigma $'s can be thought of as being composed as a difference $\sigma
_{ij}=\sigma _{i}-\sigma _{j}$, such that $\sigma _{ij}=-\sigma _{ji}$ and $%
\sigma _{ij}=\sigma _{ik}+\sigma _{kj}$. Up to now, the precise
correspondence between the unstable particles occurring in the HSG-model and
all these resonance parameters has not been worked out in the literature. So
far stable particles were associated to simple roots and unstable particles
have been identified on the quantum level as the sum of two simple roots $%
\alpha _{i},\alpha _{j}$ with $I_{ij}\neq 0$. However, these are only the
primaries and as we already mentioned, we also expect to find secondaries,
tertiaries, etc. Here we want to provide evidence that unstable particles
can in fact be related to \textbf{each} non-simple positive root, such that
the amount of unstable particles is in fact 
\begin{equation}
\text{\# of unstable particles}=\ell \,(h-2)/2.  \label{nopos}
\end{equation}
It will turn out that not all of these particles are visible in the RG
scaling function, since our bootstrap proposal will show that by
construction several of the unstable particles are unavoidably degenerate in
the mass given by (\ref{m}), such that only $\ell (\ell -1)/2$ unstable
particles will be detectable by a TBA analysis. This means whenever $h=\ell
+1$, as in $SU(\ell +1)$, we can see all unstable particles in an RG flow,
but otherwise not as for instance in $SO(2\,\ell )$.

\subsection{Bootstrap construction}

We present now three examples which we consider to be instructive to
illustrate the working of the bootstrap principle as they gradually include
new features. The $SU(3)_{2}$-HSG model contains only primary unstables, the 
$SU(4)_{2}$ model contains in addition secondaries and the $SO(8)_{2}$ model
is an example for a theory with tertiaries and mass degeneracy. Note that
for level $2$ all particles are self-conjugate.

\subsubsection{The SU(3)$_{2}$-HSG model}

Starting now with the known part of the scattering matrix (\ref{SHSG}) for
the stable particles, and leaving the remaining entries which involve
unstable particles unknown, we construct consistent solutions to the
bootstrap equations (\ref{b2}), (\ref{b3}) and (\ref{b4}). We can fix the
imaginary parts of the fusing angles by the requirement that for vanishing
resonance parameters we want to reproduce the masses predicted by the
Breit-Wigner formula. Choosing the masses of the stable particles to be $%
m_{1}=m_{2}=m$ the one for the unstable results to $m_{(12)}=\sqrt{2}m$.
This argument does not constrain the real parts of the fusing angles, such
that they are not completely fixed and still contain a certain ambiguity.
The different choices of these parameters give rise to slightly different
theories. First we consider the case $\sigma _{21}>0$.

\unitlength=1.0cm 
\begin{picture}(14.20,1.4)(-6.2,-0.70)
\put(0,0.00){\circle*{0.2}}
\put(-0.25,-0.50){$\alpha_1$}
\put(0.00,-0.01){\line(1,0){0.9}}
\put(0.35,-0.01){\line(2,1){0.4}}
\put(0.35,-0.01){\line(2,-1){0.4}}
\put(1.00,0.00){\circle*{0.2}}
\put(1.0,-0.50){$\alpha_2$}
\end{picture}

\noindent For the $SU(3)_{2}$-HSG model with this choice of the resonance
parameter, we then find the following bootstrap equations 
\begin{equation}
\tilde{S}_{l(12)}(\theta )=\tilde{S}_{l1}(\theta +(1-\nu )\sigma _{12}+i\pi
/4)\tilde{S}_{l2}(\theta -\nu \sigma _{12}-i\pi /4)
\end{equation}
from which we construct 
\begin{equation}
\tilde{S}_{SU(3)}(\theta ,\sigma _{12})=\left( 
\begin{array}{ccc}
-1 & -(\sigma _{12},2) & -((1-\nu )\sigma _{12},3) \\ 
(\sigma _{21},2) & -1 & -(\nu \sigma _{21},1) \\ 
-((\nu -1)\sigma _{12},1) & -(\nu \sigma _{12},3) & -1
\end{array}
\right) \,.  \label{SU3}
\end{equation}
Here we label the rows and columns in the order $\left\{ 1,2,(12)\right\} $.
According to the principles outlined above, the S-matrix (\ref{SU3}) allows
for the processes 
\begin{equation}
1+2\rightarrow (12),\qquad 2+(12)\rightarrow 1,\qquad (12)+1\rightarrow 2.\,
\label{pp}
\end{equation}
The related fusing angles are read off from (\ref{SU3}) as 
\begin{equation}
\eta _{12}^{(12)}=-i\pi /2+\sigma _{21},\qquad \eta _{(12)1}^{2}=-3i\pi
/4+(1-\nu )\sigma _{12},\qquad \eta _{2(12)}^{1}=-3i\pi /4+\nu \sigma _{12}\,
\end{equation}
and are interrelated through equation (\ref{eta}), which still holds even
though the $\eta $'s have non-vanishing real parts. We can employ these
fusing angles and compute the masses and decay widths by means of the
Breit-Wigner formulae (\ref{BW1}) and (\ref{BW2}). Taking again for
simplicity $m_{1}=m_{2}=m$ and in addition  $\nu =1/2$, we obtain for the
first process in (\ref{pp}) 
\begin{equation}
m_{(12)}=\sqrt{2}m\cosh \sigma _{21}/2\quad \text{and\quad }\Gamma _{(12)}=2%
\sqrt{2}m\sinh \sigma _{21}/2\,.
\end{equation}
Employing now also in the process $2+(12)\rightarrow 1$ the Breit-Wigner
formula, we construct in the limit $\sigma _{12}\rightarrow 0$ the values $%
m_{1}=m$ and $\Gamma _{1}=0$. Likewise, in the last process in (\ref{pp}) we
obtain $m_{2}=m$ and $\Gamma _{2}=0$.   

The asymptotic limit $t\rightarrow \infty $ becomes meaningful when we
operate on an energy scale at which the unstable particle has not even been
created yet, i.e.~$\Gamma _{(12)}\rightarrow \infty \equiv \sigma
_{21}\rightarrow \infty $. In that case the theory decouples into two SU(2)$%
_{2}$-models, i.e.~free Fermions, with $S_{11}=S_{22}=-1$. This is a simple
version of the decoupling rule (\ref{drule}). We consider now a different
theory with $\sigma _{12}>0$.

\unitlength=1.0cm 
\begin{picture}(14.20,1.4)(-6.2,-.70)
\put(0,0.00){\circle*{0.2}}
\put(-0.25,-0.50){$\alpha_1$}
\put(0.00,-0.01){\line(1,0){0.9}}
\put(0.6,-0.01){\line(-2,1){0.4}}
\put(0.6,-0.01){\line(-2,-1){0.4}}
\put(1.00,0.00){\circle*{0.2}}
\put(1.0,-0.50){$\alpha_2$}
\end{picture}

\noindent Taking now also in this case for simplicity $\nu =1/2$, we find
the following bootstrap satisfied 
\begin{equation}
\tilde{S}_{l(12)}(\theta )=\tilde{S}_{l2}(\theta -\sigma _{12}/2+i\pi /4)%
\tilde{S}_{l1}(\theta +\sigma _{12}/2-i\pi /4),
\end{equation}
which yields the S-matrix 
\begin{equation}
\tilde{S}_{SU(3)}(\theta ,\sigma _{21})=\left( 
\begin{array}{ccc}
-1 & (\sigma _{12},2) & -(\sigma _{12}/2,1) \\ 
-(\sigma _{21},2) & -1 & -(\sigma _{21}/2,3) \\ 
-(\sigma _{21}/2,3) & -(\sigma _{12}/2,1) & -1
\end{array}
\right) \,.  \label{SU32}
\end{equation}
The S-matrix (\ref{SU32}) allows for the processes 
\begin{equation}
2+1\rightarrow (12),\qquad 1+(12)\rightarrow 2,\qquad (12)+2\rightarrow 1,\,
\end{equation}
instead of (\ref{pp}). Now the fusing angles are read off as 
\begin{equation}
\eta _{21}^{(12)}=-i\pi /2+\sigma _{12},\qquad \eta _{1(12)}^{2}=-3i\pi
/4-\sigma _{12}/2,\qquad \eta _{2(12)}^{1}=-3i\pi /4-\sigma _{12}/2\,
\end{equation}
and also satisfy (\ref{eta}). The masses and decay width are obtained again
from (\ref{BW1}) and (\ref{BW2}) with $\sigma _{12}\rightarrow \sigma _{21}$%
. As a whole, we can think of this theory simply as being obtained from the $%
\Bbb{Z}_{2}$-Dynkin diagram automorphism which exchanges the roles of the
particles $1$ and $2$. However, since parity invariance is now broken this
is not a symmetry any more and the two theories are different. In the
asymptotic limit $\sigma _{12}\rightarrow \infty $, we obtain once again a
simple version of the decoupling rule and the theory decouples into two SU(2)%
$_{2}$-models.

We also want to make sense of joining two $SU(3)_{2}$ theories together when
one of the particles is shared, whereas the remaining ones interact
trivially. For the choice $\sigma _{21}>0$ and $\sigma _{23}>0$ this
corresponds algebraically to

\unitlength=1.0cm 
\begin{picture}(14.20,1.4)(-4.2,-0.7)
\put(0,0.00){\circle*{0.2}}
\put(-0.25,-0.50){$\alpha_1$}
\put(0.00,-0.01){\line(1,0){0.9}}
\put(0.35,-0.01){\line(2,1){0.4}}
\put(0.35,-0.01){\line(2,-1){0.4}}
\put(1.00,0.00){\circle*{0.2}}
\put(0.9,-0.50){$\alpha_2$}
\put(1.50,-0.1){$\otimes$}
\put(2.1,0.0){\circle*{0.2}}
\put(1.95,-0.50){$\alpha_2$}
\put(2.10,-0.01){\line(1,0){0.9}}
\put(2.75,-0.01){\line(-2,1){0.4}}
\put(2.75,-0.01){\line(-2,-1){0.4}}
\put(3.10,0.00){\circle*{0.2}}
\put(3.,-0.50){$\alpha_3$}
\put(3.6,-.6){\line(1,1){1.0}}
\put(4.90,0.00){\circle*{0.2}}
\put(4.8,-0.50){$\alpha_2$}
\end{picture}

\noindent Labelling the rows and columns as $\left\{ 1,2,3,(12),(23)\right\} 
$ we can simply associate the following scattering matrix to this model 
\begin{equation}
\tilde{S}_{\frac{SU(3)\otimes SU(3)}{SU(2)}}=\left( 
\begin{array}{rrrrr}
-1 & -(\sigma _{12},2) & 1 & -(\sigma _{12}/2,3) & 1 \\ 
(\sigma _{21},2) & -1 & -(\sigma _{32},2) & -(\sigma _{21}/2,1) & -(\sigma
_{23}/2,1) \\ 
1 & (\sigma _{23},2) & -1 & 1 & -(\sigma _{32}/2,3) \\ 
-(\sigma _{21}/2,1) & -(\sigma _{12}/2,3) & 1 & -1 & 1 \\ 
1 & -(\sigma _{32}/2,3) & -(\sigma _{23}/2,1) & 1 & -1
\end{array}
\right) .  \label{332}
\end{equation}
The particle $2$ is shared by the two original theories. There is a well
defined limit of this matrix, which is in agreement with the decoupling rule 
\begin{equation}
\lim_{\sigma _{21}\rightarrow \infty }\tilde{S}_{SU(3)\otimes
SU(3)/SU(2)}=\lim_{\sigma _{23}\rightarrow \infty }\tilde{S}_{SU(3)\otimes
SU(3)/SU(2)}=\tilde{S}_{SU(3)\otimes SU(2)}\,.
\end{equation}
Similarly, we can construct the remaining three cases of this kind $\sigma
_{21}<0$, $\sigma _{23}<0$ or $\sigma _{21}>0$, $\sigma _{23}<0$ or $\sigma
_{21}<0$, $\sigma _{23}>0$.

So far we have seen from this example that it is possible to develop a
consistent bootstrap and that the outcome depends on the choice of the $%
\sigma $'s. New particles are not predicted from here.

\subsubsection{The SU(4)$_{2}$-HSG model}

In comparison with the previous $SU(3)_{2}$-model the $SU(4)_{2}$-model
introduces a novel feature. Besides the fundamental unstable particles
formed from two stable particles, it also allows for the formation of
further unstable particles from the fusing of one unstable particles with a
stable one, i.e.~secondaries. We start with the case $\sigma _{21}>0,\sigma
_{23}>0$.

\unitlength=1.0cm 
\begin{picture}(14.20,1.4)(-5.8,-0.7)
\put(0,0.00){\circle*{0.2}}
\put(-0.25,-0.50){$\alpha_1$}
\put(0.00,-0.01){\line(1,0){0.9}}
\put(0.35,-0.01){\line(2,1){0.4}}
\put(0.35,-0.01){\line(2,-1){0.4}}
\put(1.00,0.00){\circle*{0.2}}
\put(0.9,-0.50){$\alpha_2$}

\put(1.0,-0.01){\line(1,0){0.9}}
\put(1.65,-0.01){\line(-2,1){0.4}}
\put(1.65,-0.01){\line(-2,-1){0.4}}
\put(2.0,0.00){\circle*{0.2}}
\put(1.9,-0.50){$\alpha_3$}

\end{picture}

\noindent Taking the two parameters of the two $SU(3)_{2}$-copies occurring
here to $\nu =1/2$, we proceed analogously as before and have the following
bootstrap equations to solve 
\begin{equation}
\begin{array}{l}
\tilde{S}_{l(12)}(\theta )=\tilde{S}_{l1}\left( \theta -\frac{\sigma _{21}}{2%
}+\frac{i\pi }{4}\right) \tilde{S}_{l2}\left( \theta +\frac{\sigma _{21}}{2}-%
\frac{i\pi }{4}\right) , \\ 
\tilde{S}_{l(23)}(\theta )=\tilde{S}_{l3}\left( \theta -\frac{\sigma _{23}}{2%
}+\frac{i\pi }{4}\right) \tilde{S}_{l2}\left( \theta +\frac{\sigma _{23}}{2}-%
\frac{i\pi }{4}\right) , \\ 
\tilde{S}_{l(123)}(\theta )=\tilde{S}_{l(12)}\left( \theta +\tilde{\mu}%
\sigma _{21}-\mu \sigma _{23}+\frac{i\pi }{4}\right) \tilde{S}_{l3}\left(
\theta +\frac{2\tilde{\mu}+1}{2}\sigma _{21}-(1+\mu )\sigma _{23}-\frac{i\pi 
}{2}\right) , \\ 
\tilde{S}_{l(123)}(\theta )=\tilde{S}_{l(23)}\left( \theta +\frac{2\tilde{\mu%
}+1}{2}\sigma _{21}-\frac{2\mu +1}{2}\sigma _{23}+\frac{i\pi }{4}\right) 
\tilde{S}_{l1}\left( \theta +\frac{2\tilde{\mu}-1}{2}\sigma _{21}-\mu \sigma
_{23}-\frac{i\pi }{2}\right) .
\end{array}
\end{equation}
Here we have already determined several fusing angles by the requirement
that the r.h.s.~of the last two equations have to be identical. The
parameters $\mu $ and $\tilde{\mu}$ enter for the same reason as in the
previous section and remain in principle free. For simplicity we choose them
now to $\mu =-\tilde{\mu}=1$. As a consistent solution to the bootstrap we
then construct 
\begin{equation}
\tilde{S}_{SU(4)}(\theta ,\sigma _{21},\sigma _{23})=\left( 
\begin{array}{cccccc}
\bullet & \bullet & \bullet & \bullet & -(\frac{\sigma _{23}-2\sigma _{21}}{2%
},3) & (-\frac{3\sigma _{21}+2\sigma _{23}}{2},2) \\ 
\bullet & \bullet & \bullet & \bullet & \bullet & -1 \\ 
\bullet & \bullet & \bullet & -(\frac{2\sigma _{32}+\sigma _{21}}{2},3) & 
\bullet & \!\!\!(-\frac{\sigma _{21}+4\sigma _{23}}{2},2)\!\!\!\! \\ 
\bullet & \bullet & \ast & \bullet & 1 & -(-\sigma _{21}-\sigma _{23},3) \\ 
\ast & \bullet & \bullet & 1 & \bullet & -(-\frac{\sigma _{21}+3\sigma _{23}%
}{2},3) \\ 
\ast & \ast & \ast & \ast & \ast & -1
\end{array}
\right) .  \label{SU4}
\end{equation}
Here we labeled the rows and columns in the order $\left\{
1,2,3,(12),(23),(123)\right\} $. The particles are all self-conjugate. We
abbreviated the entries 
\begin{equation}
\bullet \equiv \tilde{S}_{SU(3)\otimes SU(3)/SU(2)}((\sigma _{21}>0,\sigma
_{23}>0)).  \label{++}
\end{equation}
Note that (\ref{SU4}) does not completely contain the S-matrix $\tilde{S}%
_{SU(3)\otimes SU(3)/SU(2)}$, since there are some entries which do not
coincide, which indicates that the two $SU(3)$ copies still interact
non-trivially. Only the ones indicated by $\bullet $ are identical to those
in (\ref{332}). For conciseness we also did not spell out some entries
indicated by \ `$\ast $'. If the $\ast $ is in the entry $\tilde{S}_{ij}$,
we always report explicitly $\tilde{S}_{ji}$ and the omitted entry can
simply be obtained from the crossing relation (\ref{c}). In this case this
reads $(\sigma ,x)\rightarrow -(-\sigma ,4-x)$. According to the principles
outlined above, the S-matrix (\ref{SU4}) allows for the processes 
\begin{equation}
\begin{array}{rrr}
1+2\rightarrow \,\,(12),\qquad & (12)+1\rightarrow \quad \,2,\qquad & 
2+(12)\rightarrow \quad \,1,\qquad \\ 
3+2\rightarrow \,\,(23),\qquad & (23)+3\rightarrow \quad \,2,\qquad & 
2+(23)\rightarrow \quad \,3,\qquad \\ 
(12)+3\rightarrow (123),\qquad & (123)+(12)\rightarrow \quad \,3,\qquad & 
3+(123)\rightarrow (12),\qquad \\ 
(23)+1\rightarrow (123),\qquad & (123)+(23)\rightarrow \quad \,1,\qquad & 
1+(123)\rightarrow (23).\qquad
\end{array}
\label{27}
\end{equation}
The related fusing angles are read off as 
\begin{equation}
\begin{array}{lll}
\eta _{12}^{(12)}=\sigma _{21}-\frac{i\pi }{2}, & \eta _{(12)1}^{2}=-\frac{%
\sigma _{21}}{2}-\frac{3i\pi }{4}, & \eta _{2(12)}^{1}=-\frac{\sigma _{21}}{2%
}-\frac{3i\pi }{4}, \\ 
\eta _{32}^{(23)}=\sigma _{23}-\frac{i\pi }{2}, & \eta _{(23)3}^{2}=-\frac{%
\sigma _{23}}{2}-\frac{3i\pi }{4}, & \eta _{2(23)}^{3}=-\frac{\sigma _{23}}{2%
}-\frac{3i\pi }{4}, \\ 
\eta _{(12)3}^{(123)}=\frac{\sigma _{21}-2\sigma _{23}}{2}-\frac{3i\pi }{4}%
,\, & \eta _{(123)(12)}^{3}=-\sigma _{21}-\sigma _{23}-\frac{3i\pi }{4},\, & 
\eta _{3(123)}^{(12)}=\frac{\sigma _{21}+4\sigma _{23}}{2}-\frac{i\pi }{2},
\\ 
\eta _{(23)1}^{(123)}=\frac{\sigma _{23}-2\sigma _{21}}{2}-\frac{3i\pi }{4},
& \eta _{(123)(23)}^{1}=\frac{-\sigma _{21}-3\sigma _{23}}{2}-\frac{3i\pi }{4%
}, & \eta _{1(123)}^{(23)}=\frac{3\sigma _{21}}{2}+\sigma _{23}-\frac{i\pi }{%
2}.
\end{array}
\label{28}
\end{equation}
One verifies that the sum relation of the fusing angles (\ref{eta}) holds
for all possible processes. The first two lines in (\ref{27}) correspond
simply to two copies of $SU(3)_{2}$ and one does not need to comment further
on them. An interesting prediction results from the consideration of the
first two processes in the last two lines of (\ref{27}). Making in the first
process the particle $(12)$ and in the second the particle $(23)$ stable, by 
$\sigma _{2}\rightarrow \sigma _{1}$ and by $\sigma _{2}\rightarrow \sigma
_{3}$, respectively, both predict the mass of the particle $(123)$ as 
\begin{equation}
m_{(123)}\sim me^{|\sigma _{13}|/2}\;.  \label{m13}
\end{equation}
This value is precisely the one we expect from the approximation in the
Breit-Wigner formula (\ref{m}). Note that in one case we obtain $\sigma
_{13} $ and in the other $\sigma _{31}$ as a resonance parameter. The
difference results from the fact that according to the processes (\ref{27}),
the particle $(123)$ is either formed as $(1+2)+3$ or $3+(2+1)$. Thus the
different parity shows up in this process, but this has no effect on the
values for the mass. We also confirm the decoupling rule on the basis of the
constructed S-matrix 
\begin{eqnarray}
\lim_{\sigma _{21}\rightarrow \infty }\tilde{S}_{SU(4)}(\sigma _{21},\sigma
_{23}) &=&\tilde{S}_{SU(2)}+\tilde{S}_{SU(3)}(\sigma _{23})\,, \\
\lim_{\sigma _{23}\rightarrow \infty }\tilde{S}_{SU(4)}(\sigma _{21},\sigma
_{23}) &=&\tilde{S}_{SU(3)}(\sigma _{21})+\tilde{S}_{SU(2)}\,.
\end{eqnarray}

\noindent Similarly we can construct the case $\sigma _{12}>0$, $\sigma
_{32}>0$, which leads essentially to a similar qualitative behaviour and we
can therefore omit its presentation here. More interesting is the case $%
\sigma _{21}>0,\sigma _{32}>0$.

\unitlength=1.0cm 
\begin{picture}(14.20,1.4)(-5.8,-0.7)
\put(0,0.00){\circle*{0.2}}
\put(-0.25,-0.50){$\alpha_1$}
\put(0.00,-0.01){\line(1,0){0.9}}
\put(0.35,-0.01){\line(2,1){0.4}}
\put(0.35,-0.01){\line(2,-1){0.4}}
\put(1.00,0.00){\circle*{0.2}}
\put(0.9,-0.50){$\alpha_2$}

\put(1.0,-0.01){\line(1,0){0.9}}
\put(1.35,-0.01){\line(2,1){0.4}}
\put(1.35,-0.01){\line(2,-1){0.4}}
\put(2.0,0.00){\circle*{0.2}}
\put(1.9,-0.50){$\alpha_3$}

\end{picture}

\noindent Taking again the two parameters of the $SU(3)_{2}$-algebras to $%
\nu =1/2$, we proceed analogously as before and have the following bootstrap
equations to solve 
\begin{equation}
\begin{array}{l}
\tilde{S}_{l(12)}(\theta )=\tilde{S}_{l1}\left( \theta -\frac{\sigma _{21}}{2%
}+\frac{i\pi }{4}\right) \tilde{S}_{l2}\left( \theta +\frac{\sigma _{21}}{2}-%
\frac{i\pi }{4}\right) , \\ 
\tilde{S}_{l(23)}(\theta )=\tilde{S}_{l2}\left( \theta -\frac{\sigma _{32}}{2%
}+\frac{i\pi }{4}\right) \tilde{S}_{l3}\left( \theta +\frac{\sigma _{32}}{2}-%
\frac{i\pi }{4}\right) , \\ 
\tilde{S}_{l(123)}(\theta )=\tilde{S}_{l(12)}\left( \theta -\tilde{\mu}%
\sigma _{21}-\mu \sigma _{32}+\frac{i\pi }{4}\right) \tilde{S}_{l3}\left(
\theta +\frac{1-2\tilde{\mu}}{2}\sigma _{21}+(1-\mu )\sigma _{32}-\frac{i\pi 
}{2}\right) , \\ 
\tilde{S}_{l(123)}(\theta )=\tilde{S}_{l1}\left( \theta -\frac{1+2\tilde{\mu}%
}{2}\sigma _{21}-\mu \sigma _{32}-\frac{i\pi }{2}\right) \tilde{S}%
_{l(23)}\left( \theta +\frac{1-2\tilde{\mu}}{2}\sigma _{21}+\frac{1-2\mu }{2}%
\sigma _{32}-\frac{i\pi }{4}\right) .
\end{array}
\end{equation}
Again we have already fixed several fusing angles by the requirement that
the r.h.s.~of the last two equations have to be identical. The parameters $%
\mu $ and $\tilde{\mu}$ enter for the same reason as in the previous section
and remain in principle free. As a consistent solution to the bootstrap we
construct now 
\begin{equation}
\tilde{S}_{SU(4)}(\theta ,\sigma _{21},\sigma _{32})=\left( 
\begin{array}{cccccc}
\bullet & \bullet & \bullet & \bullet & -(-\frac{2\sigma _{21}+\sigma _{32}}{%
2},1) & (-\frac{(1+2\tilde{\mu})\sigma _{21}+2\mu \sigma _{32}}{2},1) \\ 
\bullet & \bullet & \bullet & \bullet & \bullet & 1 \\ 
\bullet & \bullet & \bullet & (\frac{2\sigma _{32}+\sigma _{21}}{2},3) & 
\bullet & \!\!\!(\frac{(1-2\tilde{\mu})\sigma _{21}+2(1-\mu )\sigma _{32}}{2}%
,1)\!\!\!\! \\ 
\bullet & \bullet & \ast & \bullet & (\frac{\sigma _{32}+\sigma _{21}}{2}%
,2)^{2} & (-\tilde{\mu}\sigma _{21}-\mu \sigma _{32},3) \\ 
\ast & \bullet & \bullet & (-\frac{\sigma _{32}+\sigma _{21}}{2},2)^{2} & 
\bullet & -(\frac{(1-2\tilde{\mu})\sigma _{21}+(1-2\mu )\sigma _{32}}{2},1)
\\ 
\ast & \ast & \ast & \ast & \ast & -1
\end{array}
\right) .  \label{su42}
\end{equation}
As before, we label the rows and columns in the order $\left\{
1,2,3,(12),(23),(123)\right\} $ and abbreviate the entries 
\begin{equation}
\bullet \equiv \tilde{S}_{SU(3)\otimes SU(3)/SU(2)}((\sigma _{21}>0,\sigma
_{32}>0)).\,  \label{dot}
\end{equation}
The $\ast $ entries can be obtained in the same way as in (\ref{SU4}). The
S-matrix (\ref{su42}) allows for the processes 
\begin{equation}
\begin{array}{rrr}
1+2\rightarrow \,\,(12),\qquad & (12)+1\rightarrow \quad \,2,\qquad & 
2+(12)\rightarrow \quad \,1,\qquad \\ 
2+3\rightarrow \,\,(23),\qquad & 3+(23)\rightarrow \quad \,2,\qquad & 
(23)+2\rightarrow \quad \,3,\qquad \\ 
(12)+3\rightarrow (123),\qquad & (123)+(12)\rightarrow \quad \,3,\qquad & 
3+(123)\rightarrow (12),\qquad \\ 
1+(23)\rightarrow (123),\qquad & (23)+(123)\rightarrow \quad \,1,\qquad & 
(123)+1\rightarrow (23).\qquad
\end{array}
\end{equation}
The related fusing angles are read off as 
\begin{equation}
\!\! 
\begin{array}{lll}
\eta _{12}^{(12)}=\sigma _{21}-\frac{i\pi }{2}, & \eta _{(12)1}^{2}=-\frac{%
\sigma _{21}}{2}-\frac{3i\pi }{4}, & \eta _{2(12)}^{1}=-\frac{\sigma _{21}}{2%
}-\frac{3i\pi }{4}, \\ 
\eta _{23}^{(23)}=\sigma _{32}-\frac{i\pi }{2}, & \eta _{3(23)}^{2}=-\frac{%
\sigma _{32}}{2}-\frac{3i\pi }{4}, & \eta _{(23)2}^{3}=-\frac{\sigma _{32}}{2%
}-\frac{3i\pi }{4}, \\ 
\eta _{(12)3}^{(123)}=\frac{2\sigma _{21}+4\sigma _{32}-3i\pi }{4},\, & \eta
_{(123)(12)}^{3}=-\frac{4\tilde{\mu}\sigma _{21}+4\mu \sigma _{32}+3i\pi }{4}%
,\, & \eta _{3(123)}^{(12)}=\frac{(2\mu -2)\sigma _{32}+(2\tilde{\mu}%
-1)\sigma _{21}-i\pi }{2}, \\ 
\eta _{1(23)}^{(123)}=\frac{2\sigma _{32}+4\sigma _{21}-3i\pi }{4}, & \eta
_{(23)(123)}^{1}=\frac{(4\tilde{\mu}-2)\sigma _{21}+(4\mu -2)\sigma
_{32}-3i\pi }{4}, & \eta _{(123)1}^{(23)}=-\frac{(2\tilde{\mu}+1)\sigma
_{21}+\mu \sigma _{32}+i\pi }{2}.
\end{array}
\label{66}
\end{equation}

\noindent Again one verifies that the relation (\ref{eta}) holds for all
possible processes. The first two lines in (\ref{66}) correspond again just
to two copies of $SU(3)_{2}$.\noindent\ Also in this case we predict the
mass of the particle $(123)$ to correspond to (\ref{m13}) making in the
first process the particle $(12)$ and in the second the particle $(23)$
stable, by $\sigma _{2}\rightarrow \sigma _{1}$ and by $\sigma
_{2}\rightarrow \sigma _{3}$, respectively. There are, however, some
fundamental differences with regard to the previous case. First we note that
the entries $\tilde{S}_{(12)(23)}(\theta )$ and $\tilde{S}_{(23)(12)}(\theta
)$ posses a double pole. We can interpret this as in the case when all
particles are stable in terms 
of the usual Coleman-Thun mechanism 
\FIGURE{\epsfig{file=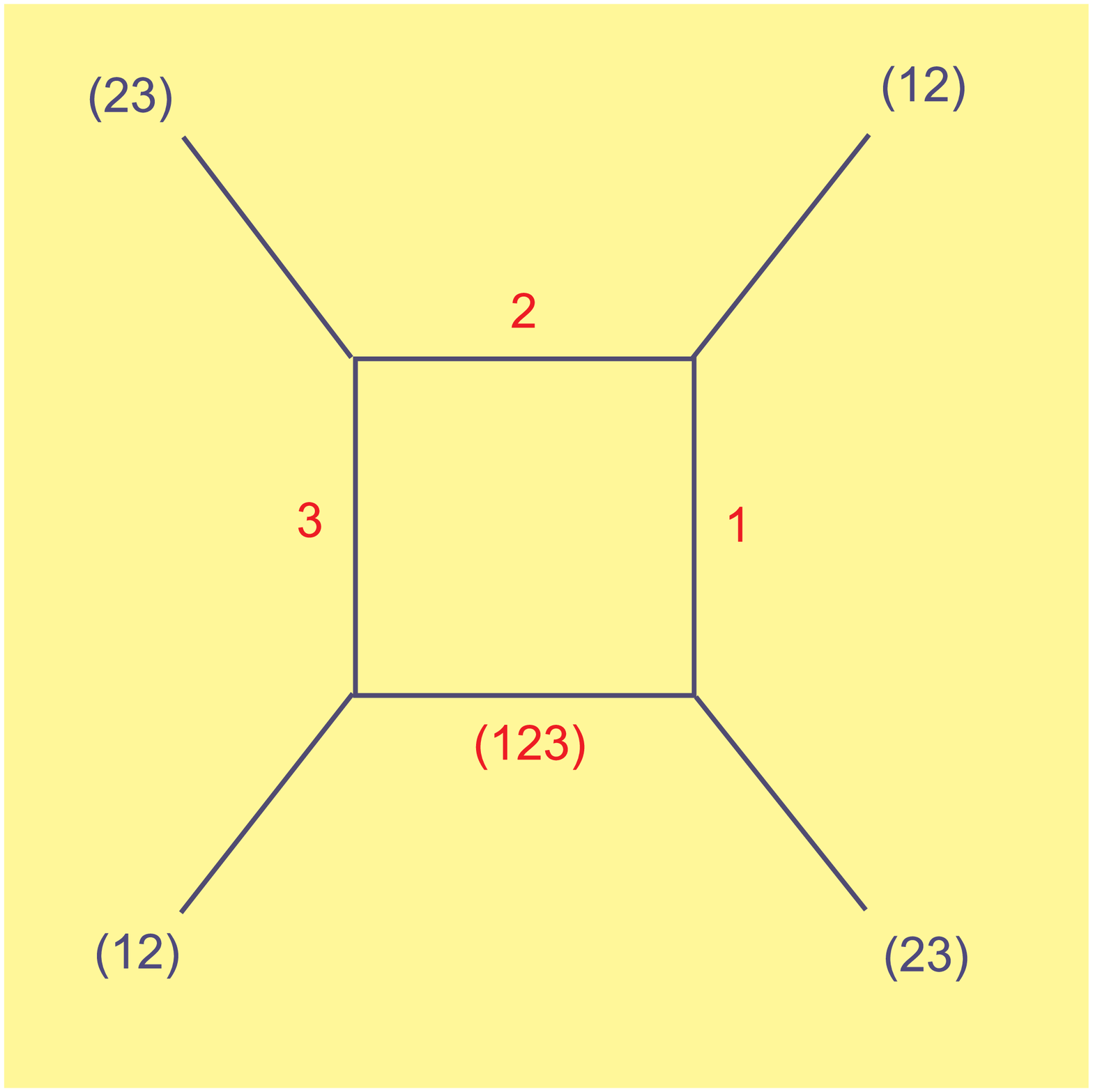,width=5.5cm} 
        \caption{Coleman-Thun mechanism.}        \label{figure2}}
\noindent \cite{CoT}. We
depict the corresponding fusing structure in figure 2. Also the decoupling
rule works differently in this case. Now we have $\sigma _{31}>\sigma
_{21},\sigma _{32}>0$ and due to the non-commutative nature of the limits we
obtain a different flow when the resonance parameters are taken to
approximate infinity. We note that apart from the $\bullet $ all the entries
tend to $\pm 1$. Thus we also confirm the decoupling rule in the version 
\begin{equation}
\lim_{\sigma _{31}\rightarrow \infty }\tilde{S}_{SU(4)}(\sigma _{21},\sigma
_{32})\sim \tilde{S}_{\frac{SU(3)\otimes SU(3)}{SU(2)}}\,.
\end{equation}
The case $\sigma _{12}>0,\sigma _{23}>0$ is very similar to this one in
behaviour and does not need to be reported. In summary, we saw in this
section that the bootstrap leads to the prediction of secondary unstable
particles with a mass computable from the Breit-Wigner formula. In addition,
we saw that by changing the order of the resonance parameters the fusing
structure does not merely change with regard to the parity, but even
develops new pole structures such as double poles.

\subsubsection{The SO(8)$_{2}$-HSG model}

\noindent For all models studied so far, we found a one-to-one
correspondence between the particle content of the theory and the amount of
resonance parameters $\sigma _{ij}$ with $i<j$ and $i,j\in \left\{ 1,\ldots
,\ell \right\} $. With regard to the RG scaling functions, which will be
computed in the next section, this means that every plateau can be
interpreted naturally as related to the onset energy for the excitation of a
particle. However, as we already indicated in section 4, this one-to-one
correspondence only holds for the $SU(\ell +1)_{2}$-HSG models. For the
remaining simply-laced algebras it turns out that the dimension of the
positive root space is always larger than the amount of parameters $\sigma
_{ij}$ at hand. We study here the simplest example which will exhibit such a
behaviour, i.e.~the $SO(8)_{2}$-HSG model. According to (\ref{nopos}) there
are 12 unstable particles, whereas the amount of available resonance
parameters is only $10$. The following analysis will show that the
generalized bootstrap equations we propose provide a satisfactory
explanation for this mismatch, in agreement with the previous understanding.
We will demonstrate that, in fact, there are $12$ particles present in the
theory, but the masses of four of them are unavoidably pairwise degenerated.
Since both the RG scaling functions and the decoupling rule are determined
by the relative mass scales of the particles in the theory, such a
degeneracy can not be unraveled in that context and only 10 particles will
be visible.

We consider now the $SO(8)_{2\text{ }}$-HSG model for the choice $\sigma
_{13}>0,\sigma _{23}>0,\sigma _{43}>0$ and label the particles as indicated
in the Dynkin diagram.

\unitlength=1.0cm 
\begin{picture}(14.20,2.4)(-5.8,-0.7)
\put(0,0.00){\circle*{0.2}}
\put(-0.25,-0.50){$\alpha_1$}
\put(0.00,-0.01){\line(1,0){0.9}}
\put(0.65,-0.01){\line(-2,1){0.4}}
\put(0.65,-0.01){\line(-2,-1){0.4}}
\put(1.00,0.00){\circle*{0.2}}
\put(0.9,-0.50){$\alpha_3$}
\put(1.02,0.){\line(0,9){0.9}}
\put(1.00,1.00){\circle*{0.2}}
\put(1.02,0.35){\line(1,2){0.2}}
\put(1.02,0.35){\line(-1,2){0.2}}
\put(1.2,0.90){$\alpha_4$}
\put(1.0,-0.01){\line(1,0){0.9}}
\put(1.35,-0.01){\line(2,1){0.4}}
\put(1.35,-0.01){\line(2,-1){0.4}}
\put(2.0,0.00){\circle*{0.2}}
\put(1.9,-0.50){$\alpha_2$}
\end{picture}

\noindent Matching every particle with a positive root of $D_{4},$ we
collect the 12 particles in the model in the set 
\begin{equation}
\mathcal{P}=\left\{
1,2,3,4,(13),(23),(43),(123),(423),(143),(1234),(12334)\right\}
\end{equation}
which allows later for very compact notations. With regard to the previous
cases we have a further novelty, namely the occurrence of tertiary unstable
particles, i.e.~$(1234)$ and $(12334)$. In addition, these two particles
exhibit the mentioned mass degeneracy, as we will see later in detail. For
the given choice of the resonance parameters we have the bootstrap equations 
\begin{equation}
\begin{array}{l}
\tilde{S}_{l(i3)}(\theta )=\tilde{S}_{l3}(\theta -\frac{\sigma _{i3}}{2}+%
\frac{i\pi }{4})\tilde{S}_{li}(\theta +\frac{\sigma _{i3}}{2}-\frac{i\pi }{4}%
)\,, \\ 
\tilde{S}_{l(ij3)}(\theta )=\tilde{S}_{lj}(\theta -\frac{\mu _{ij}\sigma
_{ji}+2\sigma _{3j}}{2}+\frac{i\pi }{2})\tilde{S}_{l(i3)}(\theta +\frac{\mu
_{ij}\sigma _{ij}+\sigma _{i3}}{2}-\frac{i\pi }{4})\,, \\ 
\tilde{S}_{l(ij3)}(\theta )=\tilde{S}_{li}(\theta -\frac{\mu _{ij}\sigma
_{ji}+2\sigma _{3i}}{2}+\frac{i\pi }{2})\tilde{S}_{l(j3)}(\theta +\frac{\mu
_{ij}\sigma _{ij}+\sigma _{j3}}{2}-\frac{i\pi }{4})\text{\thinspace }, \\ 
\tilde{S}_{l(1234)}(\theta )=\tilde{S}_{l(ki3)}(\theta -\frac{\mu
_{ki}\sigma _{ki}+\mu _{42}\sigma _{24}+\kappa }{2}+\frac{i\pi }{4})\tilde{S}%
_{lj}(\theta +\frac{\mu _{42}\sigma _{42}+\sigma _{j3}-\kappa }{2}-\frac{%
i\pi }{4})\,, \\ 
\tilde{S}_{l(12334)}(\theta )=\tilde{S}_{l(1234)}(\theta -\frac{\mu
_{42}\sigma _{42}+\sigma _{13}-\kappa +\lambda }{2}+\frac{i\pi }{4})\tilde{S}%
_{l3}(\theta -\frac{\sigma _{13}+\lambda }{2}-\frac{i\pi }{2})\,, \\ 
\tilde{S}_{l(12334)}(\theta )=\tilde{S}_{l(j3)}(\theta -\frac{\sigma
_{1j}+\lambda }{2}+\frac{i\pi }{4})\tilde{S}_{l(ki3)}(\theta +\frac{\mu
_{ki}\sigma _{ik}+\sigma _{31}-\lambda }{2}-\frac{i\pi }{2})\,,
\end{array}
\label{so8boo}
\end{equation}

\noindent The indices $i,j,k$ can take values in $\left\{ 1,2,4\right\} $,
such that no two indices are repeated and the associated particles are in $%
\mathcal{P}$. This convention will be used throughout this section. The
parameters $\mu _{ij},\kappa $ and $\lambda $ are in principle free but will
be constrained below in order to match consistently the pole structure of $%
\tilde{S}$ with the corresponding mass spectrum. The entries of the S-matrix
can be written in a very compact form, namely 
\begin{equation}
\begin{array}{l}
\tilde{S}_{i(j3)}(\theta )=(-1)^{\delta _{ij}}(\frac{\sigma _{i3}+\sigma
_{ij}}{2},1+2\delta _{i3})\,, \\ 
\tilde{S}_{i(jk3)}(\theta )=(-1)^{\delta _{ij}+\delta _{ik}}(\frac{\mu
_{jk}\sigma _{jk}+2\sigma _{i3}}{2},2)\,,\quad \qquad i\neq 3 \\ 
\tilde{S}_{(i3)(j3)}(\theta )=(-1)^{\delta _{ij}}, \\ 
\tilde{S}_{(ij3)(kl3)}(\theta )=(-1)^{\delta _{ik}\delta _{jl}}\,, \\ 
\tilde{S}_{(i3)(jk3)}(\theta )=-(-1)^{^{\delta _{ij}+\delta _{ik}}}(\frac{%
\mu _{jk}\sigma _{jk}+2\sigma _{i3}}{2},1)\,, \\ 
\tilde{S}_{(i3)(12334)}(\theta )=(\frac{\sigma _{i1}-\lambda }{2},3)\,, 
\end{array}
\end{equation}
\begin{equation}
\begin{array}{l}
\tilde{S}_{(ij3)(1234)}(\theta )=(\frac{\mu _{ij}\sigma _{ji}+\mu
_{42}\sigma _{42}-\kappa }{2},3)\,, \\ 
\tilde{S}_{(ij3)(12334)}(\theta )=-(\frac{\mu _{ij}\sigma _{ji}+\sigma
_{31}-\lambda }{2},3)\,, \\ 
\tilde{S}_{(1234)(12334)}(\theta )=(\frac{\mu _{42}\sigma _{24}+\sigma
_{31}+\kappa -\lambda }{2},3), \\ 
\tilde{S}_{(1234)(1234)}(\theta )=\tilde{S}_{(12334)(12334)}(\theta )=\tilde{%
S}_{3(jk3)}(\theta )=\tilde{S}_{(i3)(1234)}(\theta )=-1.
\end{array}
\end{equation}

\noindent According to the same principles outlined in the previous
sections, the physical scattering processes involving the fundamental
(primary) unstable particles as well as the corresponding fusing angles are 
\begin{equation}
\begin{array}{ccc}
3+j\rightarrow (j3),\quad \quad  & (j3)+3\rightarrow j, & \quad \quad
j+(j3)\rightarrow 3,
\end{array}
\end{equation}
\begin{equation}
\begin{array}{ccc}
\eta _{3j}^{(j3)}=\sigma _{j3}-\frac{i\pi }{2},\quad  & \eta _{(j3)3}^{j}=-%
\frac{\sigma _{j3}}{2}-\frac{3i\pi }{4}, & \quad \eta _{j(j3)}^{3}=-\frac{%
\sigma _{j3}}{2}-\frac{3i\pi }{4},
\end{array}
\label{pan}
\end{equation}
where we recall  $j$ can take the values $1,2$ and $4$, in correspondence
with the first equation in (\ref{so8boo}). The processes involving secondary
unstable particles are 
\begin{equation}
\begin{array}{ccc}
j+(k3)\rightarrow (jk3),\,\,\,\, & (jk3)+j\rightarrow (k3),\,\,\,\, & 
(k3)+(jk3)\rightarrow j, \\ 
k+(j3)\rightarrow (jk3),\,\,\,\, & (jk3)+k\rightarrow (j3),\,\,\,\  & 
(j3)+(jk3)\rightarrow k,
\end{array}
\label{sec}
\end{equation}
with $(jk3)\in \mathcal{P}$, such that every secondary unstable particle $%
(123),$ $(423)$ and $(143)$ can be formed in two different processes. The
associated fusing angles can also be written in a closed form 
\begin{equation}
\begin{array}{ccc}
\eta _{j(k3)}^{(jk3)}=\frac{2(\sigma _{3j}+\sigma _{jk})-3i\pi }{4},\,\, & 
\eta _{(jk3)j}^{(k3)}=\frac{\mu _{jk}\sigma _{jk}+2\sigma _{k3}-i\pi }{2}%
,\,\, & \eta _{(k3)(jk3)}^{j}=\frac{2(\mu _{jk}\sigma _{kj}+2\sigma
_{3k})-3i\pi }{4}, \\ 
\eta _{k(j3)}^{(jk3)}=\frac{2(\sigma _{3k}+\sigma _{kj})-3i\pi }{4},\,\, & 
\eta _{(jk3)k}^{(j3)}=\frac{\mu _{jk}\sigma _{jk}+2\sigma _{j3}-i\pi }{2}%
,\,\, & \eta _{(j3)(jk3)}^{k}=\frac{2(\mu _{jk}\sigma _{kj}+2\sigma
_{3j})-3i\pi }{4}.
\end{array}
\label{san}
\end{equation}
Finally, the processes which involve the tertiary unstable particles $(1234)$
and $(12334)$ are 
\begin{equation}
\begin{array}{ccc}
(kl3)+j\rightarrow (1234),\qquad  & (1234)+(kl3)\rightarrow j,\qquad  & 
j+(1234)\rightarrow (kl3), \\ 
(j3)+(kl3)\rightarrow (12334),\,\,\, & (kl3)+(12334)\rightarrow (j3),\,\,\,
& (12334)+(j3)\rightarrow (kl3), \\ 
(1234)+3\rightarrow (12334),\,\,\, & 3+(12334)\rightarrow (1234),\,\,\, & 
(12334)+(1234)\rightarrow 3,
\end{array}
\label{pro}
\end{equation}
with fusing angles

\[
\begin{array}{lll}
\eta _{(kl3)j}^{(1234)}=\frac{\mu _{kl}\sigma _{kl}+2\sigma _{j3}-i\pi }{2},
& \eta _{(1234)(kl3)}^{j}=\frac{2(\mu _{kl}\sigma _{lk}+\mu _{42}\sigma
_{42}-\kappa )-3i\pi }{4}, & \eta _{j(1234)}^{(kl3)}=\frac{2(\mu _{42}\sigma
_{24}+\kappa +2\sigma _{3j})-3i\pi }{4}
\end{array}
\]
\[
\begin{array}{lll}
\eta _{(j3)(kl3)}^{(12334)}=\frac{2(\mu _{kl}\sigma _{lk}+2\sigma
_{3j})-3i\pi }{4}, & \eta _{(kl3)(12334)}^{(j3)}=\frac{\mu _{kl}\sigma
_{kl}+\sigma _{13}+\lambda -i\pi }{2}, & \eta _{(12334)(j3)}^{(kl3)}=\frac{%
2(\sigma _{j1}-\lambda )-3i\pi }{4},
\end{array}
\]
\[
\begin{array}{lll}
\eta _{(1234)3}^{(12334)}=\frac{2(\mu _{42}\sigma _{42}-\kappa )-3i\pi }{4},
& \eta _{3(12334)}^{(1234)}=\frac{\sigma _{13}+\lambda -i\pi }{2}, & \eta
_{(12334)(1234)}^{3}=\frac{2(\mu _{42}\sigma _{24}+2\sigma _{31}-\lambda
+\kappa )-3i\pi }{4}.
\end{array}
\]
Similarly as for the $SU(4)_{2}$-HSG model studied before, one of the main
predictions, which supports the working of our proposal, are the masses of
the unstable particles. In the case of the present theory, this predictive
power is especially important, since we intend to show the degeneracy of
some of these masses. For the secondary unstable particles we can proceed as
in the previous sections and consider the processes $j+(k3)$ and $k+(j3)$
for fixed $j,k$ and make the primary unstable particles stable. By looking
at the corresponding fusing angles we find 
\begin{equation}
\left. 
\begin{array}{c}
\lim\limits_{\sigma _{3}\rightarrow \sigma _{k}}j+(k3)\rightarrow (jk3) \\ 
\lim\limits_{\sigma _{3}\rightarrow \sigma _{j}}k+(j3)\rightarrow (jk3)
\end{array}
\right\} \quad \Rightarrow \quad m_{(jk3)}\sim me^{\left| \sigma
_{jk}\right| /2},  \label{msec}
\end{equation}
that is, the masses of the secondary unstables are predicted once more
according to (\ref{m}). Most interesting is the analysis for the next
generation of unstable particles, the tertiaries.

Let us consider the formation process for the particle $(1234)$. We have to
look at the processes $(jk3)+l\rightarrow (1234)$ with $j,k,l\equiv \left\{
1,2,4\right\} $ such that $(jk3)\in \mathcal{P}$, 
\begin{equation}
\lim_{\sigma _{j}\rightarrow \sigma _{k}}\left. 
\begin{array}{c}
\lim\limits_{\sigma _{3}\rightarrow \sigma _{k}}j+(k3)\rightarrow (jk3)+l \\ 
\lim\limits_{\sigma _{3}\rightarrow \sigma _{j}}k+(j3)\rightarrow (jk3)+l
\end{array}
\right\} \rightarrow (1234)\text{ }\Rightarrow m_{(1234)}\sim \sqrt{2}me^{%
\frac{|\sigma _{l3}|}{2}}.  \label{m11}
\end{equation}
This equation requires some comments. Reading (\ref{m11}) for each
individual value of $l$, we end up with the condition $\sigma _{k}=\sigma
_{j}=\sigma _{3}$ for $j,k\in \left\{ 1,2,4\right\} \neq l$. This should
hold for all values of $l$, such that when we demand the uniqueness of the
mass $m_{(1234)}$ we end up with the condition $\sigma _{1}=\sigma
_{2}=\sigma _{3}=\sigma _{4}$. This means 
\begin{equation}
m_{(1234)}\sim \sqrt{2}m\,.
\end{equation}
and the particle $(1234)$ will also be invisible in any RG flow. Let us now
consider the tertiary particle $(12334).$ The processes to be investigated
are the ones listed above in (\ref{pro}) 
\begin{eqnarray}
\lim_{\sigma _{j}\rightarrow \sigma _{k}}\left. 
\begin{array}{c}
\lim\limits_{\sigma _{3}\rightarrow \sigma _{k}}j+(k3)\rightarrow (jk3)+l \\ 
\lim\limits_{\sigma _{3}\rightarrow \sigma _{j}}k+(j3)\rightarrow (jk3)+l
\end{array}
\right\}  &\rightarrow &\stackunder{\sigma _{l}\rightarrow \sigma _{3}}{%
(1234)}+3\rightarrow (12334)\Rightarrow m_{(12334)}\sim m,  \label{m22} \\
\lim_{\sigma _{l}\rightarrow \sigma _{3}}(l3)+\left. 
\begin{array}{c}
\lim\limits_{\sigma _{3}\rightarrow \sigma _{k}}j+(k3) \\ 
\lim\limits_{\sigma _{3}\rightarrow \sigma _{j}}k+(j3)
\end{array}
\right\}  &\rightarrow &(l3)+\stackunder{\sigma _{j}\rightarrow \sigma _{k}}{%
(jk3)}\rightarrow (12334)\Rightarrow m_{(12334)}\sim m.\quad   \label{m12}
\end{eqnarray}
We presented here the complete chain of processes leading to the formation
of the unstable particle $(12334).$ As can be seen from (\ref{m12}) and (\ref
{m22}), making all intermediate particles stable amounts to the equality of
all resonance parameters, so that there is no $\sigma $-dependence in the
final expression for the mass. Therefore, with regard to the mass spectrum,
the particle $(12334)$ is indistinguishable from the stable particles in the
theory.

Due to the fact that the formation of $(1234)$ and $(12334)$ requires all
resonance parameters to be equal, the parameters $\kappa $ and $\lambda $
can never be fixed by the mass analysis. Similarly the parameters $\mu _{ij}$
can also not be constrained further by means of the mass analysis. However,
there is still a further check to be carried out, namely the decoupling
rule. For the choice mentioned $\sigma _{13}>0,\sigma _{23}>0,\sigma _{43}>0$
we expect to find the following decoupling flow 
\begin{eqnarray}
&&\lim_{\sigma _{43},\sigma _{23},\sigma _{13}\rightarrow \infty }\tilde{S}%
_{SO(8)}(\sigma _{13},\sigma _{23},\sigma _{43})=\lim_{\sigma _{43},\sigma
_{23}\rightarrow \infty }\tilde{S}_{SU(4)}(\sigma _{23},\sigma _{43})+\tilde{%
S}_{SU(2)}  \nonumber \\
&=&\lim_{\sigma _{43}\rightarrow \infty }\tilde{S}_{SU(3)}(\sigma _{43})+%
\tilde{S}_{SU(2)}+\tilde{S}_{SU(2)}=\tilde{S}_{SU(2)}+\tilde{S}_{SU(2)}+%
\tilde{S}_{SU(2)}.  \label{dso8}
\end{eqnarray}
In order to achieve consistency of $\tilde{S}$ with (\ref{dso8}) the
following constraints have to be satisfied 
\begin{eqnarray}
\left| \mu _{ij}\right| &>&2,  \nonumber \\
\quad \left| \kappa _{1}\right| &>&2,\quad \left| \kappa _{2}\right|
>0,\quad \left| \kappa _{4}\right| >0,  \nonumber \\
\left| \lambda _{2}\right| &>&1,\quad \left| \lambda _{4}\right| >1,\quad
\left| \lambda _{1}\right| >0,  \nonumber \\
\left| \kappa _{1}+\mu _{12}\right| &>&0,\quad \left| \kappa _{1}+\mu
_{14}\right| >0,  \nonumber \\
\left| \kappa _{2}+\mu _{42}\right| &>&2,\quad \left| \kappa _{2}-\mu
_{12}\right| >0,  \label{con} \\
\left| \kappa _{4}+\mu _{42}\right| &>&0,\quad \left| \kappa _{4}-\mu
_{42}\right| >2,  \nonumber \\
\left| \lambda _{1}-\kappa _{1}+1\right| &>&0,\quad \left| \lambda _{1}+\mu
_{14}+1\right| >0,\quad \left| \lambda _{1}+\mu _{12}+1\right| >0,\quad 
\nonumber \\
\left| \lambda _{2}-\kappa _{2}-\mu _{42}\right| &>&0,\quad \left| \lambda
_{2}-\mu _{12}\right| >0,\quad \left| \lambda _{2}-\mu _{42}\right| >0, 
\nonumber \\
\left| \lambda _{4}-\kappa _{4}+\mu _{42}\right| &>&0,\quad \left| \lambda
_{4}-\mu _{42}\right| >0,\quad \left| \lambda _{4}-\mu _{14}\right| >0, 
\nonumber
\end{eqnarray}
for 
\begin{equation}
\kappa =\kappa _{1}\sigma _{13}+\kappa _{2}\sigma _{23}+\kappa _{4}\sigma
_{43\quad }\mathrm{and}\quad \lambda =\lambda _{1}\sigma _{13}+\lambda
_{2}\sigma _{23}+\lambda _{4}\sigma _{43}.  \label{kl}
\end{equation}
A possible, but not unique, choice which allows to satisfy the conditions (%
\ref{con}) is 
\begin{equation}
\mu _{ij}=\kappa _{1}=\lambda _{1}=5/2,\quad \lambda _{2}=\lambda
_{4}=3/2\quad \mathrm{and\quad }\kappa _{4}=\kappa _{2}=7/2.
\end{equation}
We summarize the main new observations from this example: Tertiary unstable
particles can be formed, but in this case they are unavoidably mass
degenerate in comparison with a primary and a stable particle.

\subsection{Thermodynamic Bethe ansatz analysis}

In this section we want to compare the findings of the previous section with
an alternative method. We employ for this the thermodynamic Bethe ansatz 
\cite{TBAZam} in order to compute the scaling function in dependence of the
inverse temperature. One should stress that the only input into this
approach is the proper S-matrix involving stable particles and not $\tilde{S}
$. As discussed in \cite{CFKM} for the first time for the HSG-models, we
expect to find the typical staircase pattern, in which each step can be
related to the energy scale of the formation of at least an unstable
particle. The central equations to solve in this context are the
TBA-equations 
\begin{equation}
rm_{a}^{i}\cosh \theta =\varepsilon _{a}^{i}(\theta
)+\sum\limits_{b=1}^{\ell }\sum\limits_{j=1}^{\tilde{\ell}}\Phi
_{ab}^{ij}\ast L_{b}^{j}(\theta )\quad   \label{ptba}
\end{equation}
By the symbol '$\ast $' we denote the rapidity convolution of two functions
defined as $f\ast g(\theta ):=\int d\theta ^{\prime }/2\pi \,f(\theta
-\theta ^{\prime })g(\theta ^{\prime })$. The renormalization group
parameter $r$ is related to the temperature $T$ as $r=m/T$, where $m$ is the
mass scale of the lightest particle. We also re-defined the masses as usual
by $m_{a}^{i}\rightarrow m_{a}^{i}/m$. The pseudo-energies $\varepsilon
_{a}^{i}(\theta )$ are related to the functions $L_{b}^{j}(\theta )=\ln
(1+e^{-\varepsilon _{b}^{j}(\theta )})$. The kernels in the integrals carry
the information of the dynamical interaction of the system and are given by 
\begin{equation}
\Phi _{ab}^{ij}(\theta )=-i\frac{d}{d\theta }\ln S_{ab}^{ij}(\theta )\,.
\label{kernel}
\end{equation}

\noindent Recall that in general we characterize the particles by two
quantum numbers $(a,i)$ as explained in section 3, unlike in the last
sections where we could drop one as we were dealing with level 2 only. For
instance for the \textbf{\~{g}}$_{2}$-HSG scattering matrix (\ref{SHSG}) the
kernel simply reads 
\begin{equation}
\Phi ^{ij}(\theta )=\tilde{I}_{ij}\cosh ^{-1}(\theta +\sigma _{ij})\,,
\end{equation}
with $\tilde{I}$ \ being the incidence matrix of \textbf{\~{g}}. Having
solved the equations (\ref{ptba}) for the pseudo-energies $\varepsilon
_{i}(\theta )$ one can compute the scaling function 
\begin{equation}
c(r)=\frac{3\,r}{\pi ^{2}}\sum_{i,a}m_{a}^{i}\int\limits_{0}^{\infty
}d\theta \,\cosh \theta \,(L_{a}^{i}(\theta )+L_{a}^{i}(-\theta ))\,.
\label{cr}
\end{equation}
Due to its non-linear nature the TBA equations are known not to be solvable
analytically, albeit in certain regions analytical approximations exist. We
shall now solve these equations numerically. According to the above
discussion we are particularly interested in higher rank Lie algebras, since
the decoupling rule (\ref{drule}) will be most complex in that case. Even
though solving (\ref{ptba}) numerically is a relatively simple iteration
problem, its full solution for high ranks requires a considerable
computational effort. In order to tackle such situations a sophisticated
procedure has been developed \cite{Jul}, which allows to compute in a
reasonable short time high rank scaling functions to a very high precision.

However, in many cases one can use some standard techniques, see e.g.~\cite
{TBAZam,CFKM}, and approximate (\ref{ptba}) by the so-called constant
TBA-equations and predict the values for $c$. In a large regime for $\theta $%
, when $r$ is small, one may approximate $\varepsilon _{a}^{i}(\theta
)=\varepsilon _{a}^{i}=$ $const.$ By standard TBA arguments \cite{TBAZam} it
follows that the effective central charge is expressible as 
\begin{equation}
c_{\text{eff}}=\frac{6}{\pi ^{2}}\sum\limits_{a=1}^{\ell }\sum\limits_{i=1}^{%
\tilde{\ell}}\mathcal{L}\left( \frac{x_{a}^{i}}{1+x_{a}^{i}}\right) ,\qquad
x_{a}^{i}=\prod\limits_{b=1}^{\ell }\prod\limits_{j=1}^{\tilde{\ell}%
}(1+x_{b}^{j})^{N_{ab}^{ij}}\,\,  \label{ceff}
\end{equation}
with $\mathcal{L}(x)=\sum_{n=1}^{\infty }x^{n}/n^{2}+\ln x\ln (1-x)/2$
denoting Rogers dilogarithm, $x_{a}^{i}=\exp (-\varepsilon _{a}^{i})$ and $%
N_{ab}^{ij}=1/2\pi \int_{-\infty }^{\infty }d\theta \,\,\Phi
_{ab}^{ij}(\theta )\,$. We will exploit this fact below.

\subsubsection{The SU(4)$_{2}$-HSG model}

As already discussed as an example in sections 3.1 and 3.2.2, we have two
qualitatively different behaviours for this case. For the ordering $\sigma
_{21}>0,\sigma _{23}>0$ we expect to find 3 plateaux in the scaling
function, whereas the ordering $\sigma _{21}>0,\sigma _{32}>0$ is associated
to a different decoupling rule with 4 plateaux. This is the behaviour the
decoupling rule predicts and which we already saw confirmed when taking the
limits of the scattering matrix obtained from our bootstrap proposal. In
addition we now also find this behaviour being validated with a TBA
analysis. Figure 3 contains the numerical outcome of the solutions of the
equations (\ref{ptba}), (\ref{cr}) and agrees with our expectations. Note
that we do not only predict correctly the height of the plateaux, but also
their onset at $\ln (r/2)\sim \sigma /2$, which is in agreement with the
approximation to the Breit-Wigner formula (\ref{m}). From this analysis we
can now support our reasoning with regard to the comment on the occurrence
of the absolute value in the Breit-Wigner formula. The two cases presented
in figure 3 only differ by the signs of the resonance parameter $\sigma _{12}
$ and are evidently not identical. In one case we have $\sigma _{13}=\sigma
_{12}+\sigma _{23}=90$ leading to an additional plateau and in the other we
have $\sigma _{13}=\sigma _{12}+\sigma _{23}=30$, which does not give an
onset related to $\sigma _{13}$ as the decoupling between the vertices $1$
and $2$ already took place.

\FIGURE{\epsfig{file=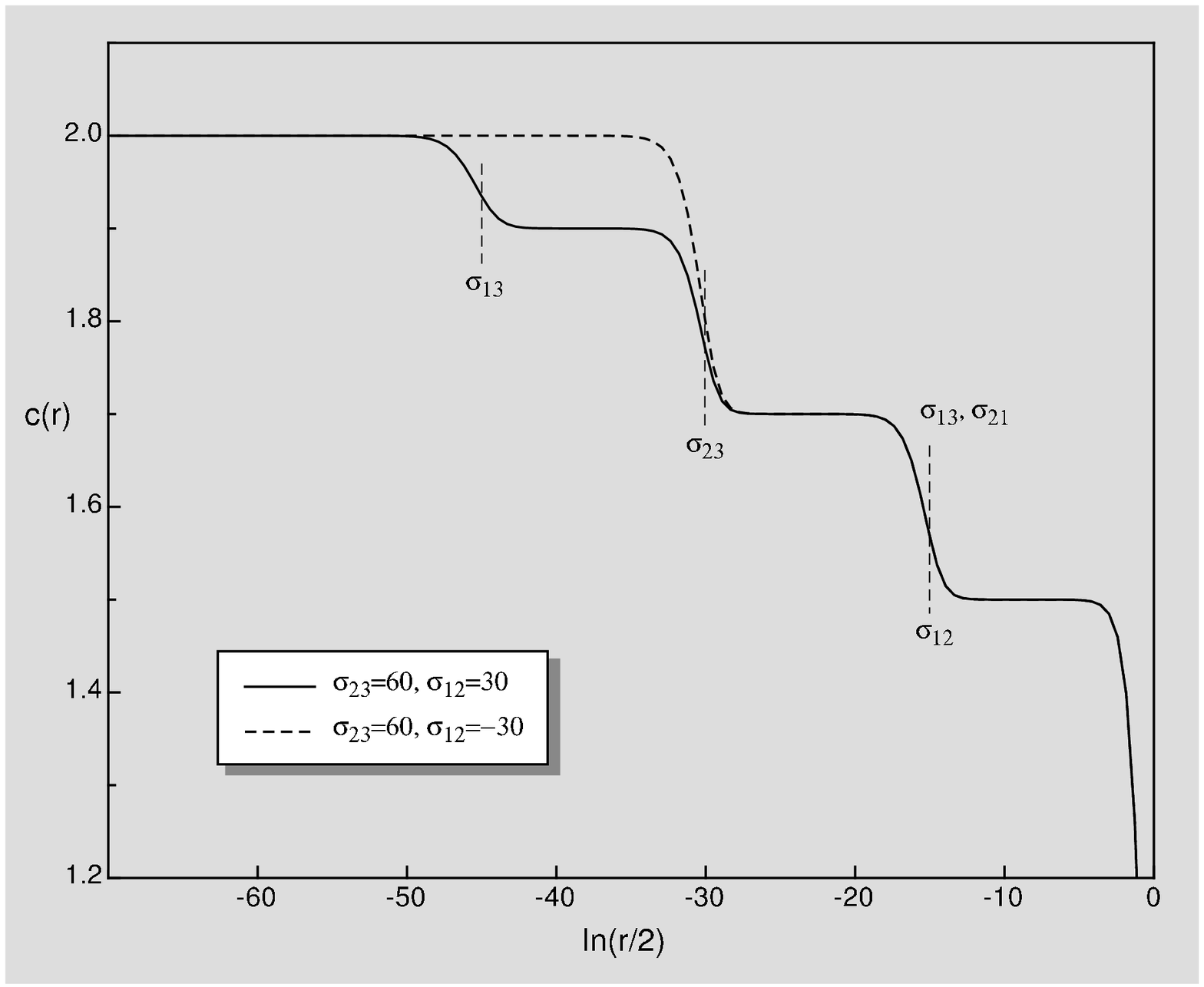,width=85mm} 
     \caption{The $SU(4)_2$ RG flow.}        \label{figure3}}

\noindent These findings seem to contradict 
some result previously obtained from a
form factor analysis \cite{CF8}. In there as in all previous publications on
the subject, e.g.~\cite{HSGS,CFKM}, an absolute value was used in the
Breit-Wigner formula, such that the signs of the resonance parameters were
handled quite casual. In fact, the results presented in figure 1 in \cite
{CF8} for $SU(4)_{2}$ correspond to the values $\sigma _{12}=50$, $\sigma
_{23}=-20$ rather than $\sigma _{12}=50$, $\sigma _{23}=20$ which agrees
with the decoupling rule. Likewise matters are clarified below for the $%
SU(5)_{2}$-HSG case.

\FIGURE{\epsfig{file=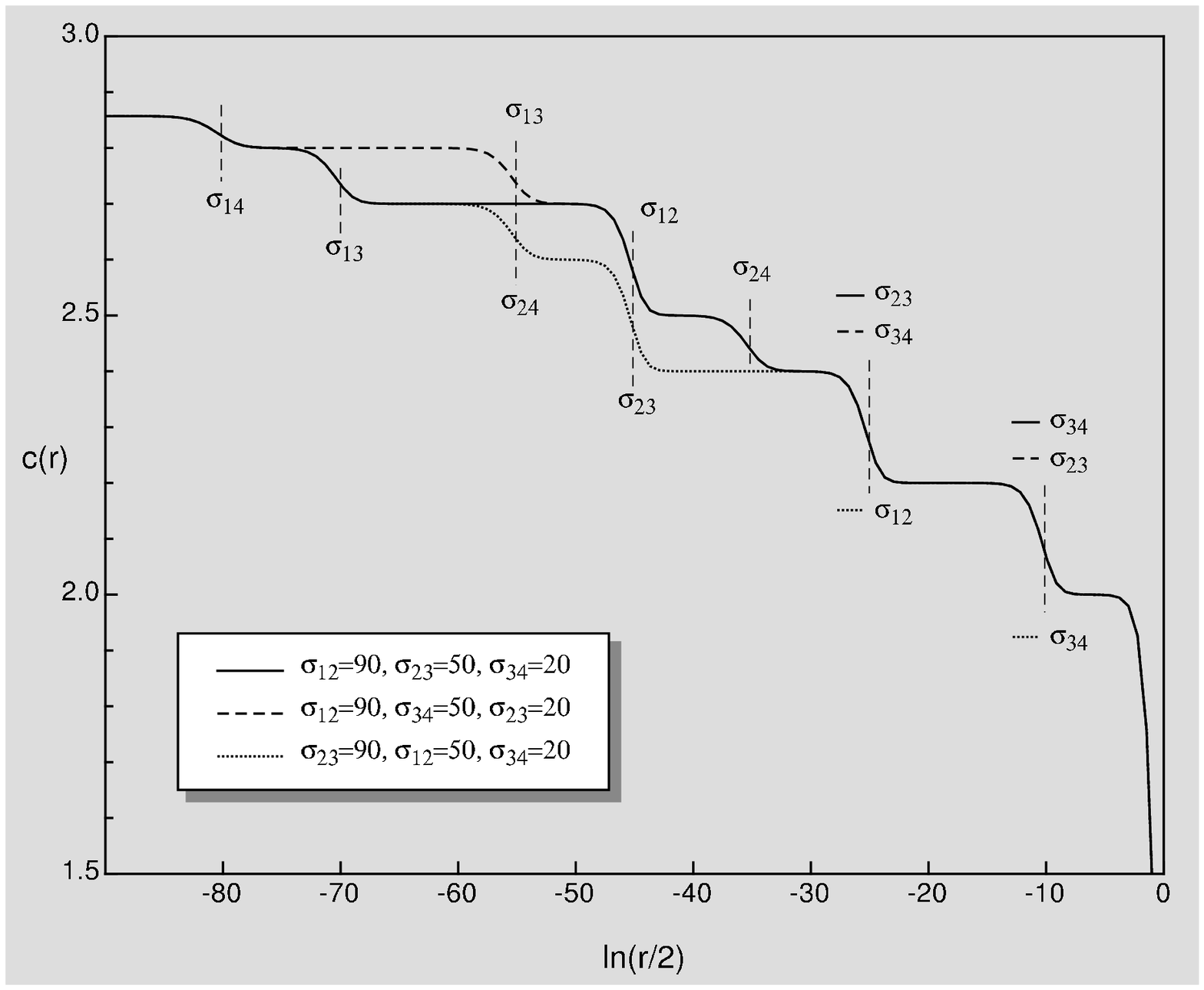,width=85mm} 
      \caption{The $SU(5)_2$ RG flow.}        \label{figure4}}

\noindent The occurrence 
of additional plate- aux which could not be associated to
primary unstable particles for several algebras was noted in \cite{Jul}.
Their occurrence in the $SU(4)_{2}$, $SU(5)_{2}$, $SO(8)_{2}$-HSG models, as
will also be discussed in the next section, was also commented upon in \cite
{ALoch}. This latter observation was only ba- sed on a TBA-calculation of the
finite size scaling function. In \cite{ALoch} the statement was made that
these further plateaux could possibly be related to ``new'' unstable
particles in the spectrum. However, predictive rules for the fixed points of
the RG flow such as the decoupling rule (\ref{drule}) and also for the
on-set as the bootstrap principle, which we both provide here, were absent.
\subsubsection{The SU(5)$_{2}$-HSG model}

This case already allows for many more possibilities, that is,
potentially $6!$ different orderings of the resonance parameters. Let us
discuss one case in more detail and present two more in a shorter form. We
choose the values for the resonance parameters directly associated to the
links of the Dynkin diagram, i.e.~the primary unstables, as $\sigma _{23}=90$%
, $\sigma _{12}=50$, $\sigma _{34}=20$, such that the remaining parameters
result to $\sigma _{14}=160$, $\sigma _{13}=140$, $\sigma _{24}=110$.
According to the decoupling rule we can then predict the renormalization
group flow. We report here the flow from the ultraviolet to the infrared,
indicating the resonance parameters responsible for the particular onsets
together with the Dynkin diagrams of the new underlying algebra and their
related Lie groups and Virasoro central charges:

\unitlength=1.0cm 
\begin{picture}(15.20,7.5)(-1.0,-5.8)
\put(0,1.00){\circle*{0.2}}
\put(-0.25,.50){$\alpha_1$}
\put(0.00,.99){\line(1,0){0.9}}
\put(1.00,1.00){\circle*{0.2}}
\put(0.9,0.50){$\alpha_2$}
\put(1.00,0.99){\line(1,0){0.9}}
\put(2.00,1.00){\circle*{0.2}}
\put(1.9,0.50){$\alpha_3$}
\put(2.00,0.99){\line(1,0){0.9}}
\put(3.00,1.00){\circle*{0.2}}
\put(2.9,0.50){$\alpha_4$}
\put(9.3,0.85){\footnotesize $ SU(5)$}
\put(11.8,0.8){$\frac{20}{7} \sim 2.86$}
\put(-1.5,0.00){$\rightarrow \sigma_{14}$}
\put(0,0.00){\circle*{0.2}}
\put(-0.25,-0.50){$\alpha_1$}
\put(0.00,-0.01){\line(1,0){0.9}}
\put(1.00,0.00){\circle*{0.2}}
\put(0.9,-0.50){$\alpha_2$}
\put(1.00,-0.01){\line(1,0){0.9}}
\put(2.00,0.00){\circle*{0.2}}
\put(1.9,-0.50){$\alpha_3$}
\put(2.50,-0.1){$\otimes$}
\put(3.1,0.0){\circle*{0.2}}
\put(2.95,-0.50){$\alpha_2$}
\put(3.10,-0.01){\line(1,0){0.9}}
\put(4.10,0.00){\circle*{0.2}}
\put(4.,-0.50){$\alpha_3$}
\put(4.10,-0.01){\line(1,0){0.9}}
\put(5.10,0.00){\circle*{0.2}}
\put(5.,-0.50){$\alpha_4$}
\put(5.6,-.6){\line(1,1){1.0}}
\put(6.90,0.00){\circle*{0.2}}
\put(6.8,-0.50){$\alpha_2$}
\put(7.10,-0.01){\line(1,0){0.9}}
\put(8.10,0.00){\circle*{0.2}}
\put(8.,-0.50){$\alpha_3$}
\put(9.3,-0.15){$ \frac{SU(4)^2}{SU(3)}$}
\put(12.3,-0.2){$2.8$}
\put(-1.5,-1.00){$\rightarrow\sigma_{13}$}
\put(0,-1.00){\circle*{0.2}}
\put(-0.25,-1.50){$\alpha_1$}
\put(0.00,-1.01){\line(1,0){0.9}}
\put(1.00,-1.00){\circle*{0.2}}
\put(0.9,-1.50){$\alpha_2$}
\put(1.50,-1.1){$\otimes$}
\put(2.1,-1.0){\circle*{0.2}}
\put(1.95,-1.50){$\alpha_2$}
\put(2.10,-1.01){\line(1,0){0.9}}
\put(3.10,-1.00){\circle*{0.2}}
\put(3.,-1.50){$\alpha_3$}
\put(3.10,-1.01){\line(1,0){0.9}}
\put(4.10,-1.00){\circle*{0.2}}
\put(4.,-1.50){$\alpha_4$}
\put(4.6,-1.6){\line(1,1){1.0}}
\put(5.90,-1.00){\circle*{0.2}}
\put(5.8,-1.50){$\alpha_2$}
\put(9.3,-1.15){$ \frac{SU(4) \otimes SU(3) }{SU(2)}$}
\put(12.3,-1.2){$2.7$}
\put(-1.5,-2.00){$\rightarrow\sigma_{24}$}
\put(0,-2.00){\circle*{0.2}}
\put(-0.25,-2.50){$\alpha_1$}
\put(0.00,-2.01){\line(1,0){0.9}}
\put(1.00,-2.00){\circle*{0.2}}
\put(0.9,-2.50){$\alpha_2$}
\put(1.50,-2.1){$\otimes$}
\put(2.1,-2.0){\circle*{0.2}}
\put(1.95,-2.50){$\alpha_2$}
\put(2.10,-2.01){\line(1,0){0.9}}
\put(3.10,-2.00){\circle*{0.2}}
\put(3.,-2.50){$\alpha_3$}
\put(3.60,-2.1){$\otimes$}
\put(4.3,-2.0){\circle*{0.2}}
\put(4.15,-2.50){$\alpha_3$}
\put(4.30,-2.01){\line(1,0){0.9}}
\put(5.30,-2.00){\circle*{0.2}}
\put(5.2,-2.50){$\alpha_4$}
\put(5.8,-2.6){\line(1,1){1.0}}
\put(7.0,-2.00){\circle*{0.2}}
\put(6.9,-2.50){$\alpha_2$}
\put(7.60,-2.1){$\otimes$}
\put(8.30,-2.00){\circle*{0.2}}
\put(8.2,-2.50){$\alpha_3$}
\put(9.3,-2.15){$ \frac{SU(3)^{\otimes 3} }{SU(2)^{\otimes 2}}$}
\put(12.3,-2.2){$2.6$}
\put(-1.5,-3.00){$\rightarrow\sigma_{23}$}
\put(0,-3.00){\circle*{0.2}}
\put(-0.25,-3.50){$\alpha_1$}
\put(0.00,-3.01){\line(1,0){0.9}}
\put(1.00,-3.00){\circle*{0.2}}
\put(0.9,-3.50){$\alpha_2$}
\put(1.50,-3.1){$\otimes$}
\put(2.1,-3.0){\circle*{0.2}}
\put(1.95,-3.50){$\alpha_3$}
\put(2.10,-3.01){\line(1,0){0.9}}
\put(3.10,-3.00){\circle*{0.2}}
\put(3.,-3.50){$\alpha_4$}
\put(9.3,-3.15){\footnotesize $ SU(3)^{\otimes 2} $}
\put(12.3,-3.2){$2.4$}
\put(-1.5,-4.00){$\rightarrow\sigma_{12}$}
\put(0,-4.00){\circle*{0.2}}
\put(-0.25,-4.50){$\alpha_1$}
\put(.40,-4.1){$\otimes$}
\put(1.00,-4.00){\circle*{0.2}}
\put(0.9,-4.50){$\alpha_2$}
\put(1.50,-4.1){$\otimes$}
\put(2.1,-4.0){\circle*{0.2}}
\put(1.95,-4.50){$\alpha_3$}
\put(2.10,-4.01){\line(1,0){0.9}}
\put(3.10,-4.00){\circle*{0.2}}
\put(3.,-4.50){$\alpha_4$}
\put(9.3,-4.15){\footnotesize $ SU(3) \otimes SU(2)^{\otimes 2} $}
\put(12.3,-4.2){$2.2$}
\put(-1.5,-5.00){$\rightarrow\sigma_{34}$}
\put(0,-5.00){\circle*{0.2}}
\put(-0.25,-5.50){$\alpha_1$}
\put(.40,-5.1){$\otimes$}
\put(1.00,-5.00){\circle*{0.2}}
\put(0.9,-5.50){$\alpha_2$}
\put(1.50,-5.1){$\otimes$}
\put(2.1,-5.0){\circle*{0.2}}
\put(1.95,-5.50){$\alpha_3$}
\put(2.50,-5.1){$\otimes$}
\put(3.10,-5.00){\circle*{0.2}}
\put(3.,-5.50){$\alpha_4$}
\put(9.3,-5.15){\footnotesize $ SU(2)^{\otimes 4} $}
\put(12.3,-5.2){$2$}
\end{picture}
\noindent Comparing this computation with the numerical outcome of our TBA
analysis, depicted in figure 4, \noindent reproduces very well this flow in
form of the dotted line together with the onset at $\ln (r/2)\sim \sigma /2$
for each resonance parameter. Note the explicit occurrence of the secondary
unstable particles. The other two examples presented in figure 4 are 
\begin{equation}
\begin{array}{lll}
& SU(5) & 20/7\sim 2.86 \\ 
\rightarrow \sigma _{14}=160/\sigma _{14}=160\qquad \qquad & SU(4)^{\otimes
2}/SU(3) & 2.8 \\ 
\rightarrow \sigma _{13}=140/\sigma _{13}=110 & SU(4)\otimes SU(3)/SU(2) & 
2.7 \\ 
\rightarrow \sigma _{12}=90/\sigma _{12}=90 & SU(4)\otimes SU(2) & 2.5 \\ 
\rightarrow \sigma _{24}=70/\sigma _{24}=70 & SU(3)^{\otimes 2}\otimes
SU(2)/SU(2)\qquad \quad & 2.4 \\ 
\rightarrow \sigma _{23}=50/\sigma _{34}=50 & SU(3)\otimes SU(2)^{\otimes 2}
& 2.2 \\ 
\rightarrow \sigma _{34}=20/\sigma _{23}=20 & SU(2)^{\otimes 4} & 2
\end{array}
\label{44}
\end{equation}
Note that both orderings in (\ref{44}) give rise to the same flows which is
once again in agreement with the decoupling rule. The difference between
these two orderings is, however, that we are dealing with theories which
contain different mass spectra of the unstable particles, such that the
onset varies in both cases according to (\ref{m}).

\subsubsection{The SO(8)$_{2}$-HSG model}

Being a rank 4 algebra as the previous example, also in this case we have
potentially $6!$ different orderings of the resonance parameters. Let us
study this case for the ordering 
\begin{equation}
\sigma _{24}=140>\sigma _{23}=90>\sigma _{21}=\sigma _{14}=70>\sigma
_{34}=50>\sigma _{13}=20\,.
\end{equation}
Then, according to the decoupling rule, we predict the flow

\unitlength=1.0cm 
\begin{picture}(15.20,8.5)(-1.0,-6.0)
\put(0,1.00){\circle*{0.2}}
\put(-0.25,.50){$\alpha_1$}
\put(0.00,.99){\line(1,0){0.9}}
\put(1.00,1.00){\circle*{0.2}}
\put(0.9,0.50){$\alpha_3$}
\put(1.00,0.99){\line(1,0){0.9}}
\put(2.00,1.00){\circle*{0.2}}
\put(1.9,0.50){$\alpha_4$}
\put(1.02,0.99){\line(0,1){0.9}}
\put(1.00,1.99){\circle*{0.2}}
\put(1.2,1.8){$\alpha_2$}
\put(9.3,0.85){\footnotesize $ SO(8)$}
\put(12.3,0.8){$3$}
\put(-1.5,0.00){$\rightarrow\sigma_{24}$}
\put(0,0.00){\circle*{0.2}}
\put(-0.25,-0.50){$\alpha_1$}
\put(0.00,-0.01){\line(1,0){0.9}}
\put(1.00,0.00){\circle*{0.2}}
\put(0.9,-0.50){$\alpha_3$}
\put(1.00,-0.01){\line(1,0){0.9}}
\put(2.00,0.00){\circle*{0.2}}
\put(1.9,-0.50){$\alpha_2$}
\put(2.50,-0.1){$\otimes$}
\put(3.1,0.0){\circle*{0.2}}
\put(2.95,-0.50){$\alpha_1$}
\put(3.10,-0.01){\line(1,0){0.9}}
\put(4.10,0.00){\circle*{0.2}}
\put(4.,-0.50){$\alpha_3$}
\put(4.10,-0.01){\line(1,0){0.9}}
\put(5.10,0.00){\circle*{0.2}}
\put(5.,-0.50){$\alpha_4$}
\put(5.6,-.6){\line(1,1){1.0}}
\put(6.90,0.00){\circle*{0.2}}
\put(6.8,-0.50){$\alpha_1$}
\put(7.10,-0.01){\line(1,0){0.9}}
\put(8.10,0.00){\circle*{0.2}}
\put(8.,-0.50){$\alpha_3$}
\put(9.3,-0.15){$ \frac{SU(4)^2}{SU(3)}$}
\put(12.3,-0.2){$2.8$}
\put(-1.5,-1.00){$\rightarrow\sigma_{23}$}
\put(.00,-1.00){\circle*{0.2}}
\put(-0.1,-1.50){$\alpha_2$}
\put(.50,-1.1){$\otimes$}
\put(1.1,-1.0){\circle*{0.2}}
\put(0.95,-1.50){$\alpha_1$}
\put(1.10,-1.01){\line(1,0){0.9}}
\put(2.10,-1.00){\circle*{0.2}}
\put(2.,-1.50){$\alpha_3$}
\put(2.10,-1.01){\line(1,0){0.9}}
\put(3.10,-1.00){\circle*{0.2}}
\put(3.,-1.50){$\alpha_4$}
\put(9.3,-1.15){\footnotesize $ SU(4) \otimes SU(2) $} 
\put(12.3,-1.2){$2.5$}
\put(-1.5,-2.00){$\rightarrow\sigma_{14}$}
\put(0.00,-2.00){\circle*{0.2}}
\put(-0.1,-2.50){$\alpha_2$}
\put(.50,-2.1){$\otimes$}
\put(1.1,-2.0){\circle*{0.2}}
\put(0.95,-2.50){$\alpha_1$}
\put(1.10,-2.01){\line(1,0){0.9}}
\put(2.10,-2.00){\circle*{0.2}}
\put(2.,-2.50){$\alpha_3$}
\put(2.60,-2.1){$\otimes$}
\put(3.3,-2.0){\circle*{0.2}}
\put(3.15,-2.50){$\alpha_3$}
\put(3.30,-2.01){\line(1,0){0.9}}
\put(4.30,-2.00){\circle*{0.2}}
\put(4.2,-2.50){$\alpha_4$}
\put(4.8,-2.6){\line(1,1){1.0}}
\put(6.0,-2.00){\circle*{0.2}}
\put(5.9,-2.50){$\alpha_3$}
\put(9.3,-2.15){$ \frac{SU(3)^{\otimes 2} \otimes SU(2) }{SU(2)}$}
\put(12.3,-2.2){$2.4$}
\put(-1.5,-3.00){$\rightarrow\sigma_{21}$}
\put(.0,-3.2){is already decoupled}
\put(-1.5,-4.00){$\rightarrow\sigma_{34}$}
\put(0,-4.00){\circle*{0.2}}
\put(-0.25,-4.50){$\alpha_2$}
\put(.40,-4.1){$\otimes$}
\put(1.00,-4.00){\circle*{0.2}}
\put(0.9,-4.50){$\alpha_4$}
\put(1.50,-4.1){$\otimes$}
\put(2.1,-4.0){\circle*{0.2}}
\put(1.95,-4.50){$\alpha_1$}
\put(2.10,-4.01){\line(1,0){0.9}}
\put(3.10,-4.00){\circle*{0.2}}
\put(3.,-4.50){$\alpha_3$}
\put(9.3,-4.15){\footnotesize $ SU(3) \otimes SU(2)^{\otimes 2} $}
\put(12.3,-4.2){$2.2$}
\put(-1.5,-5.00){$\rightarrow\sigma_{13}$}
\put(0,-5.00){\circle*{0.2}}
\put(-0.25,-5.50){$\alpha_1$}
\put(.40,-5.1){$\otimes$}
\put(1.00,-5.00){\circle*{0.2}}
\put(0.9,-5.50){$\alpha_2$}
\put(1.50,-5.1){$\otimes$}
\put(2.1,-5.0){\circle*{0.2}}
\put(1.95,-5.50){$\alpha_3$}
\put(2.50,-5.1){$\otimes$}
\put(3.10,-5.00){\circle*{0.2}}
\put(3.,-5.50){$\alpha_4$}
\put(9.3,-5.15){\footnotesize $ SU(2)^{\otimes 4} $}
\put(12.3,-5.2){$2$}
\end{picture}

\FIGURE{\epsfig{file=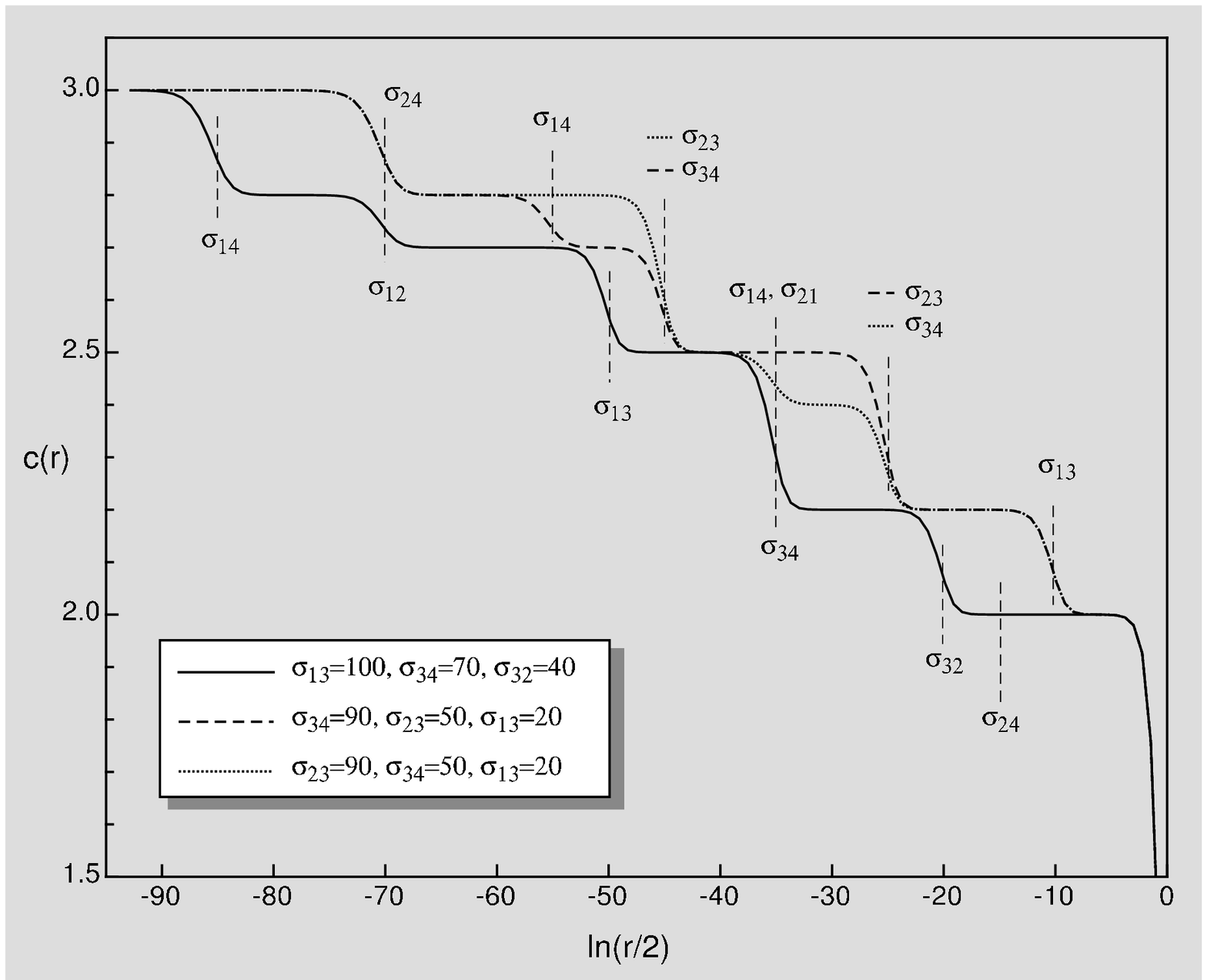,width=85mm} 
       \caption{The $SO(8)_2$ RG flow.}        \label{figure5}}

\noindent From the decoupling at $\sigma _{14}$ we observe that it is
important to keep track of the simple root labels on the vertices. The
cancellation will be different whether one decouples next with $\sigma _{13}$
or $\sigma _{34}$. The presented values for the RG fixed points may be
compared with the dotted line in figure 5, which is again the numerical
outcome of a TBA-analysis. We present two more flows in figure 5. Once again
the two different orderings lead to the same flows, but the onsets vary
according to (\ref{m}). An important fact to notice from all graphs is that
we only observe the primary and secondary unstables in this analysis and the
tertiaries remain invisible. This is in agreement with our bootstrap
analysis in section 3.2.3, which provides a consistent explanation of the
claim that all positive roots but the simple ones should be related to
unstable particles. 
\begin{equation}
\begin{array}{lll}
& SO(8) & 3 \\ 
\rightarrow \sigma _{14}=170/\sigma _{24}=140\qquad \qquad  & SU(4)^{\otimes
2}/SU(3) & 2.8 \\ 
\rightarrow \sigma _{12}=140/\sigma _{14}=110 & SU(4)\otimes
SU(3)/SU(2)\qquad \qquad  & 2.7 \\ 
\rightarrow \sigma _{13}=100/\sigma _{34}=90 & SU(4)\otimes SU(2) & 2.5 \\ 
\rightarrow \sigma _{34}=70/\sigma _{23}=50 & SU(2)^{\otimes 2}\otimes SU(3)
& 2.2 \\ 
\rightarrow \ast \ast \ast \ast \ast /\sigma _{21}=30 & \text{is already
decoupled} & 
\end{array}
\end{equation}

\begin{equation}
\begin{array}{lll}
\rightarrow \sigma _{32}=40/\sigma _{13}=20 & SU(2)^{\otimes 4} & 2 \\ 
\rightarrow \sigma _{24}=30/\ast \ast \ast \ast \ast \ast & \text{is already
decoupled} & 
\end{array}
\nonumber
\end{equation}

In summary, we conclude from this section that the number of unstable
particles is actually 12, in agreement with (\ref{nopos}), but for the
mentioned reason only 10 can be seen in the RG flow.

\subsubsection{The (E$_{6}$)$_{2}$-HSG model}

In this case we have already $15!$ different possible orderings. We want to
present two more explicit examples, which will be instructive since in
comparison with the previous ones they add more non-trivial structure,
namely mass degeneracy. Note that this degeneracy is not the unavoidable one
of the tertiary unstables as discussed in the previous section, but it
arises through the particular choice of the primary resonance parameters.
This is nonetheless instructive as we will demonstrate that the decoupling
rule also works well in that case. We label the particles as indicated in
the following Dynkin diagram:

\unitlength=1.cm 
\begin{picture}(14.20,2.2)(-4.0,0.2)
\put(0,1.00){\circle*{0.2}}
\put(-0.25,.50){$\alpha_1$}
\put(0.00,.99){\line(1,0){0.9}}
\put(1.00,1.00){\circle*{0.2}}
\put(0.9,0.50){$\alpha_3$}
\put(1.00,0.99){\line(1,0){0.9}}
\put(2.00,1.00){\circle*{0.2}}
\put(1.9,0.50){$\alpha_4$}
\put(2.02,0.99){\line(0,1){0.9}}
\put(2.00,1.99){\circle*{0.2}}
\put(2.2,1.8){$\alpha_2$}
\put(2.00,0.99){\line(1,0){0.9}}
\put(3.00,1.00){\circle*{0.2}}
\put(2.9,0.50){$\alpha_5$}
\put(3.00,0.99){\line(1,0){0.9}}
\put(4.00,1.00){\circle*{0.2}}
\put(3.9,0.50){$\alpha_6$}
\end{picture}

\noindent We choose once again first the ordering for the primary resonance
parameters 
\begin{equation}
\sigma _{13}=100>\sigma _{34}=80>\sigma _{45}=60>\sigma _{56}=40>\sigma
_{24}=20\,.
\end{equation}
According to the decoupling rule we predict then the flow:

\begin{equation}
\begin{array}{lll}
& E_{6} & \frac{36}{7}\sim 5.\,\allowbreak 14 \\ 
\rightarrow \sigma _{16}=280 & SO(10)^{\otimes 2}/SO(8) & 5 \\ 
\rightarrow \sigma _{15}=240 & SO(10)\otimes SU(5)/SU(4) & \frac{34}{7}\sim
4.\,\allowbreak 86 \\ 
\rightarrow \sigma _{14}=\sigma _{36}=180\quad & SO(8)\otimes SU(5)\otimes
SU(3)/SU(4)\otimes SU(2) & \frac{319}{70}\sim 4.\,\allowbreak 56 \\ 
\rightarrow \sigma _{12}=160 & \text{is already decoupled} &  \\ 
\rightarrow \sigma _{35}=140 & SU(5)\otimes SU(4)\otimes SU(3)/SU(3)\otimes
SU(2) & \frac{61}{14}\sim 4.\,\allowbreak 36 \\ 
\rightarrow \sigma _{26}=120 & SU(4)^{\otimes 3}\otimes SU(3)/SU(3)^{\otimes
2}\otimes SU(2) & 4.3 \\ 
\rightarrow \sigma _{13}=\sigma _{46}=100 & SU(4)^{\otimes 2}\otimes
SU(3)\otimes SU(2)/SU(3)\otimes SU(2)\quad & 4 \\ 
\rightarrow \sigma _{25}=\sigma _{34}=80 & SU(3)^{\otimes 3}\otimes
SU(2)^{\otimes 2}/SU(2)^{\otimes 2} & 3.6 \\ 
\rightarrow \sigma _{32}=\sigma _{45}=60 & SU(3)^{\otimes 2}\otimes
SU(2)^{\otimes 2} & 3.4 \\ 
\rightarrow \sigma _{56}=40 & SU(3)\otimes SU(2)^{\otimes 4} & 3.2 \\ 
\rightarrow \sigma _{24}=20 & SU(2)^{\otimes 6} & 3
\end{array}
\end{equation}

\noindent These analytical predictions are well confirmed with the outcome
of our TBA-analysis as presented by the solid line in figure 6. Note that
eight particles are pairwise degenerate and we therefore expect to find $%
15-8/2=11$ plateaux in the flow. The first step which corresponds to one of
these degeneracies occurs for instance at $\sigma _{14}=\sigma _{36}$ and we
have to apply the decoupling rule twice at this point.

\FIGURE{\epsfig{file=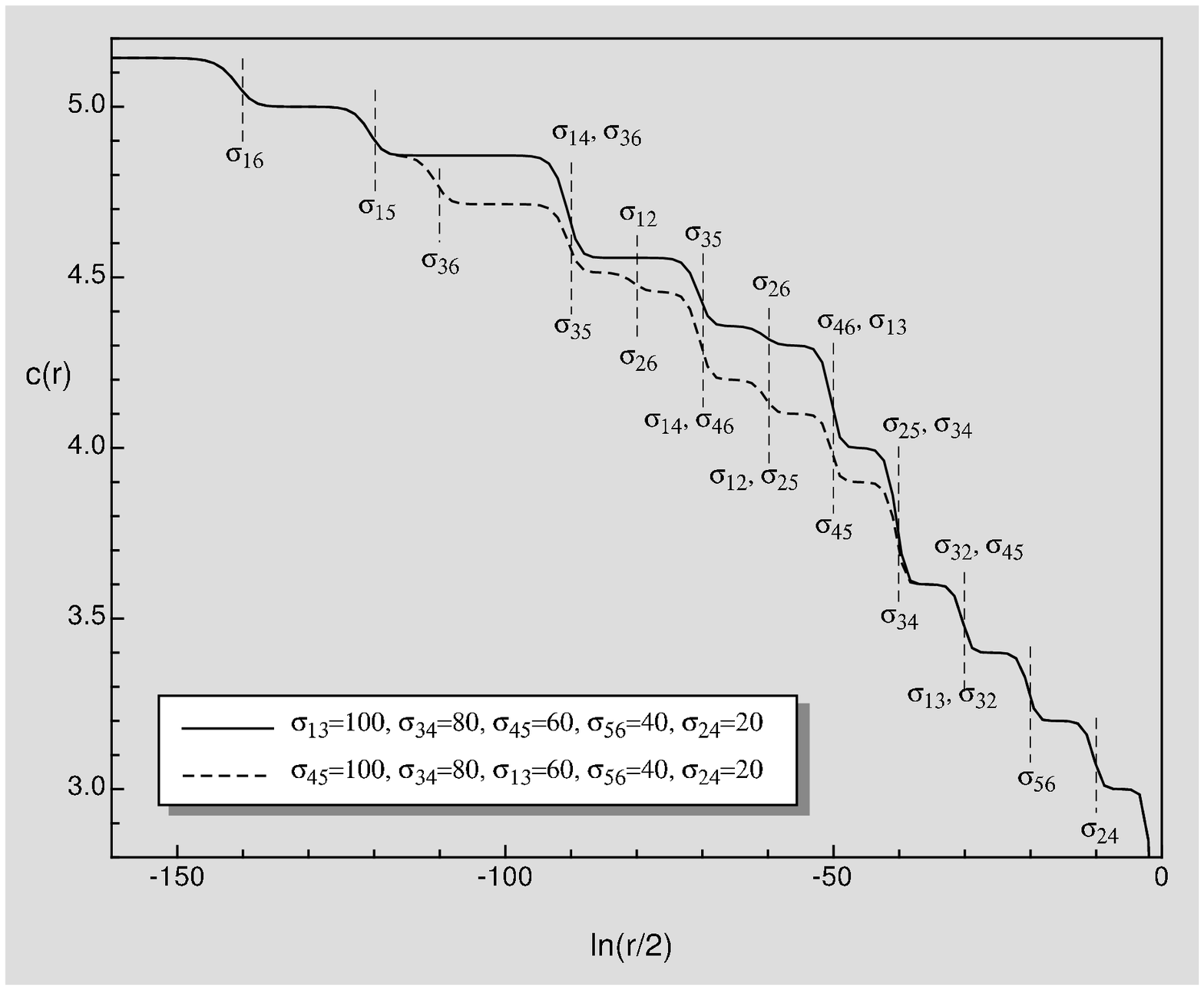,width=88mm} 
       \caption{The $(E_{6})_2$ RG flow. }        \label{figure6}}

\noindent The dashed line corresponds to the flow 
\begin{equation}
\begin{array}{lll}
& E_{6} & \frac{36}{7}\sim 5.\,\allowbreak 14 \\ 
\rightarrow \sigma _{16}=280 & SO(10)^{\otimes 2}/SO(8) & 5 \\ 
\rightarrow \sigma _{15}=240 & SO(10)\otimes SU(5)/SU(4) & \frac{34}{7}\sim
4.\,\allowbreak 86 \\ 
\rightarrow \sigma _{36}=220 & SU(5)^{\otimes 2}\otimes SO(8)/SU(4)^{\otimes
2} & \frac{33}{7}\sim 4.\,\allowbreak 71 \\ 
\rightarrow \sigma _{35}=180 & SU(5)^{\otimes 2}/SU(3) & \frac{158}{35}\sim
4.\,\allowbreak 51 \\ 
\rightarrow \sigma _{26}=160 & SU(4)^{\otimes 2}\otimes SU(5)/SU(3)^{\otimes
2} & \frac{156}{35}\sim 4.\,\allowbreak 46 \\ 
\rightarrow \sigma _{14}=\sigma _{46}=140\quad & SU(4)^{\otimes 2}\otimes
SU(3)^{\otimes 2}/SU(2)^{\otimes 2}\otimes SU(3)\quad & 4.2 \\ 
\rightarrow \sigma _{12}=\sigma _{25}=120 & SU(4)\otimes SU(3)^{\otimes
3}/SU(2)^{\otimes 3} & 4.1 \\ 
\rightarrow \sigma _{45}=100 & SU(4)\otimes SU(3)^{\otimes 2}/SU(2) & 3.9 \\ 
\rightarrow \sigma _{34}=80 & SU(3)^{\otimes 3} & 3.6 \\ 
\rightarrow \sigma _{13}=\sigma _{32}=60 & SU(3)^{\otimes 2}\otimes
SU(2)^{\otimes 2} & 3.4 \\ 
\rightarrow \sigma _{56}=40 & SU(3)\otimes SU(2)^{\otimes 4} & 3.2 \\ 
\rightarrow \sigma _{24}=20 & SU(2)^{\otimes 6} & 3
\end{array}
\end{equation}
Now only six particles are pairwise degenerate and we expect to find $%
15-6/2=12$ plateaux, which is precisely what we see in figure 6. Overall we
have seen in this section that even for this involved case the decoupling
rule is confirmed.

\subsubsection{The SU(4)$_{3}$-HSG model}

In order to support the working of the decoupling rule also for higher level
algebras we will now consider one level 3 example, i.e.~$SU(4)_{3}$. In this
model we have six particles. When labeling the rows and columns in the order
\{$(1,1)$, $(1,2)$, $(1,3)$, $(2,1)$, $(2,2)$, $(2,3)$\}, the scattering
matrix in this case reads 
\begin{equation}
S=\left( 
\begin{array}{cccccc}
\lbrack 2]_{0}^{-1} & e^{-\frac{i\pi }{3}}[1]_{\sigma _{12}} & 1 & 
-[1]_{0}^{-1} & e^{-\frac{2\pi i}{3}}[2]_{\sigma _{12}} & 1 \\ 
e^{\frac{i\pi }{3}}[1]_{\sigma _{21}} & [2]_{0}^{-1} & e^{-\frac{i\pi }{3}%
}[1]_{\sigma _{23}} & e^{\frac{2\pi i}{3}}[2]_{\sigma _{21}} & -[1]_{0}^{-1}
& e^{-\frac{2\pi i}{3}}[2]_{\sigma _{23}} \\ 
1 & e^{\frac{i\pi }{3}}[1]_{\sigma _{32}} & [2]_{0}^{-1} & 1 & e^{\frac{2\pi
i}{3}}[2]_{\sigma _{32}} & -[1]_{0}^{-1} \\ 
-[1]_{0}^{-1} & e^{-\frac{2\pi i}{3}}[2]_{\sigma _{12}} & 1 & 
[2]_{0}^{-2}[4]_{0}^{-1} & -e^{-\frac{i\pi }{3}}[1]_{\sigma _{12}} & 1 \\ 
e^{\frac{2\pi i}{3}}[2]_{\sigma _{21}} & -[1]_{0}^{-1} & e^{-\frac{2\pi i}{3}%
}[2]_{\sigma _{23}} & -e^{\frac{i\pi }{3}}[1]_{\sigma _{21}} & 
[2]_{0}^{-2}[4]_{0}^{-1} & -e^{-\frac{i\pi }{3}}[1]_{\sigma _{23}} \\ 
1 & e^{\frac{2\pi i}{3}}[2]_{\sigma _{32}} & -[1]_{0}^{-1} & 1 & -e^{\frac{%
i\pi }{3}}[1]_{\sigma _{32}} & [2]_{0}^{-2}[4]_{0}^{-1}
\end{array}
\right)  \label{su4}
\end{equation}
where we abbreviated $\left[ x\right] _{\sigma }:=\sinh \frac{1}{2}\left(
\theta +\sigma -\frac{i\pi x}{3}\right) /\sinh \frac{1}{2}\left( \theta
+\sigma +\frac{i\pi x}{3}\right) $. As in the previous cases we can now
solve the TBA equations numerically and according to the fusing rule we
expect as in the level 2 case either 3 or 4 plateaux depending on the
ordering $\sigma _{21}>0,\sigma _{23}>0$ or $\sigma _{21}>0,\sigma _{32}>0$,
respectively. 
\begin{equation}
\begin{array}{lll}
& SU(4)_{3} & \frac{24}{7}\sim 3.\,\allowbreak 49 \\ 
\rightarrow \sigma _{23}=60\qquad \qquad \qquad & SU(2)_{3}\otimes
SU(3)_{3}\qquad \qquad & 2.8\qquad \qquad \qquad \qquad \\ 
\rightarrow \sigma _{21}=30 & \left( SU(2)_{3}\right) ^{\otimes 3} & 2.4 \\ 
\mathrm{or} &  &  \\ 
& SU(4)_{3} & \frac{24}{7}\sim 3.\,\allowbreak 49 \\ 
\rightarrow \sigma _{13}=90\qquad & \left( SU(3)_{3}\right) ^{\otimes
2}/SU(2)_{3}\qquad \qquad & 3.2 \\ 
\rightarrow \sigma _{23}=60 & SU(2)_{3}\otimes SU(3)_{3} & 2.8\qquad \qquad
\qquad \\ 
\rightarrow \sigma _{12}=30 & \left( SU(2)_{3}\right) ^{\otimes 3} & 2.4
\end{array}
\end{equation}

\noindent Indeed figure 7 confirms this behaviour in form of the solid and
dashed line. Unfortuna- 
\FIGURE{\epsfig{file=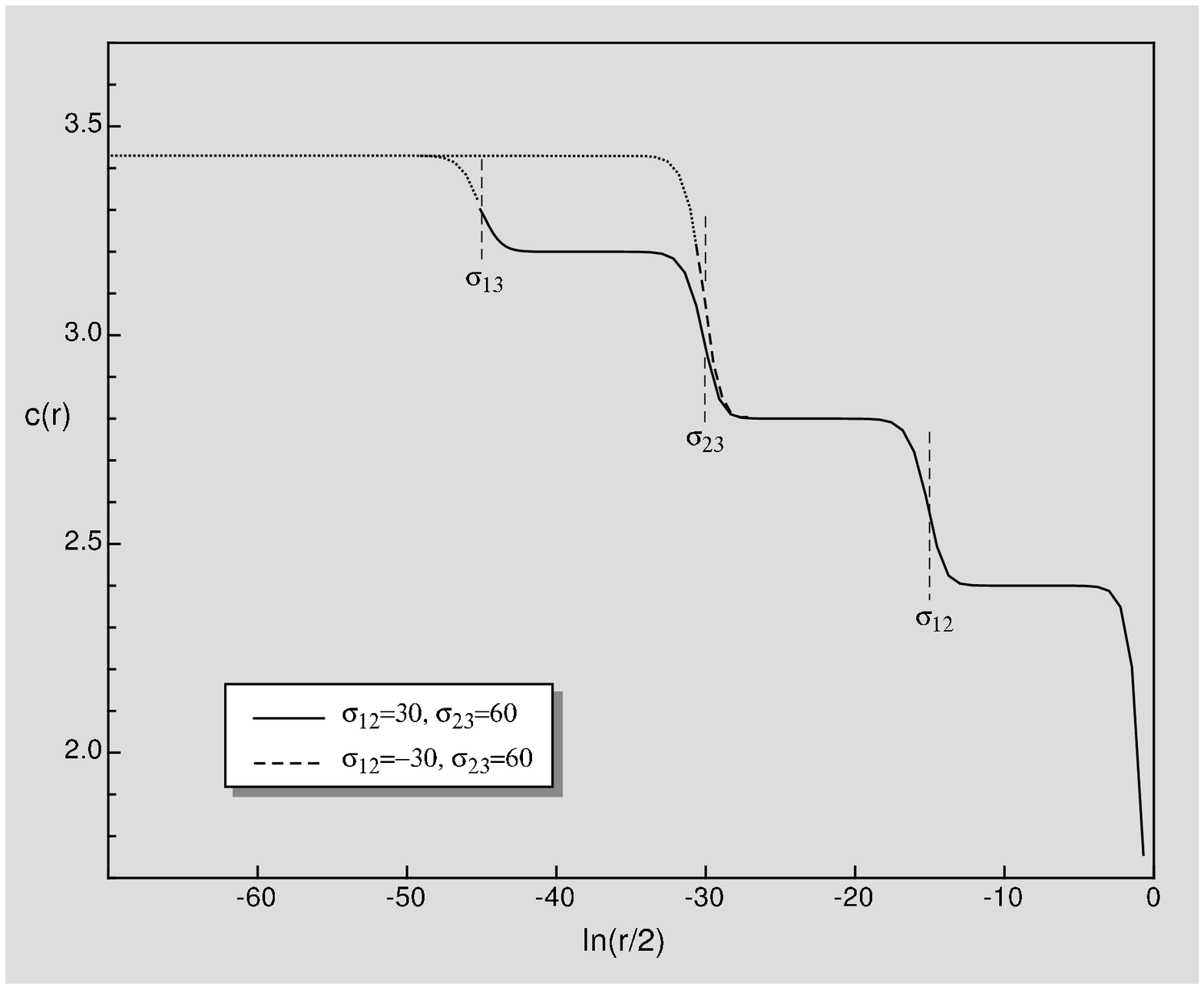,width=85mm} 
       \caption{The $SU(4)_3$ RG flow.}        \label{figure7}} \noindent
tely, the iterative procedure used for solving the TBA equations ceases to
converge when we reach the value ln$(r/2)$ equal to the highest resonance
parameter in both cases. None- theless, the highest plateau may be computed
analytically simply from the constant TBA equations (\ref{ceff}). In figure
7 we indicate these analytical values, which we can not obtain numerically
at present, by the dotted lines. The other RG fixed points may be computed
similarly. We evaluate\bigskip \bigskip

\begin{eqnarray}
x_{1}^{1} &=&x_{2}^{1}=x_{1}^{3}=x_{2}^{1}=\frac{\sin (\frac{2\pi }{7})\sin (%
\frac{4\pi }{7})}{\sin (\frac{\pi }{7})\sin (\frac{5\pi }{7})}-1,\quad
x_{1}^{2}=x_{2}^{2}=\frac{\sin ^{2}(\frac{3\pi }{7})}{\sin (\frac{\pi }{7}%
)\sin (\frac{5\pi }{7})}-1, \\
c_{\text{eff}} &=&\frac{6}{\pi ^{2}}\left[ 4\mathcal{L}\left( \frac{x_{1}^{1}%
}{1+x_{1}^{1}}\right) +2\mathcal{L}\left( \frac{x_{2}^{2}}{1+x_{2}^{2}}%
\right) \right] \,=\frac{24}{7}\sim 3.43\,.
\end{eqnarray}
The important fact here is the confirmation of the decoupling rule for
higher levels, sustained by the occurrence of the additional plateau at 3.2.

\subsubsection{The $Sp(4)_{2}$-HSG model}

Having provided some evidence for the working of the decoupling rule (\ref
{drule}) at higher levels, we now want to extend this also to non-simply
laced algebras, for which consistent S-matrices were proposed in \cite{CK}.
Since a TBA analysis has not been carried out for the non-simply laced case,
the following will also provide support for the working of the proposal in 
\cite{CK}. According to the latter the $Sp(4)_{2}$-HSG model is comprised
out of four stable particles 
\begin{equation}
\mathcal{P}=\left\{ (1,1)=\overline{(1,1)},\text{ }(1,2)=\overline{(3,2)},%
\text{ }(2,2)=\overline{(2,2)},\text{ }(3,2)=\overline{(1,2)}\right\} ,
\label{pso5}
\end{equation}
where, as before, we refer to every particle by two quantum numbers $(a,i)$.
Labeling the rows and columns of the S-matrix in the same order as in (\ref
{pso5}) and particularizing the closed formulae provided in \cite{CK} yields 
\begin{equation}
S_{Sp(4)_{2}}=\left( 
\begin{array}{cccc}
-1 & -(\sigma _{12},2) & \left( \sigma _{12},1\right) \left( \sigma
_{12},3\right)  & -(\sigma _{12},2) \\ 
-(\sigma _{21},2) & i(0,-2) & -\left( 0,-1\right) \left( 0,-3\right)  & 
-i(0,-2) \\ 
\left( \sigma _{21},1\right) \left( \sigma _{21},3\right)  & -\left(
0,-1\right) \left( 0,-3\right)  & (0,-2)^{2} & -\left( 0,-1\right) \left(
0,-3\right)  \\ 
-(\sigma _{21},2) & -i(0,-2) & -\left( 0,-1\right) \left( 0,-3\right)  & 
i(0,-2)
\end{array}
\right) \,,  \label{so5}
\end{equation}
where $\sigma _{12}=-$ $\sigma _{21}$ is the only resonance parameter in the
theory and we employed the building blocks (\ref{building}). Carrying out
the logarithmic derivative of the individual components of $S_{Sp(4)_{2}}$
yields the kernel 
\begin{equation}
\Phi _{Sp(4)_{2}}=\left( 
\begin{array}{cccc}
0 & \frac{1}{\cosh (\theta +\sigma _{12})} & \frac{2\sqrt{2}\cosh (\theta
+\sigma _{12})}{\cosh 2(\theta +\sigma _{12})} & \frac{1}{\cosh (\theta
+\sigma _{12})} \\ 
\frac{1}{\cosh (\theta +\sigma _{21})} & \frac{-1}{\cosh \theta } & \frac{-2%
\sqrt{2}\cosh \theta }{\cosh 2\theta } & \frac{-1}{\cosh \theta } \\ 
\frac{2\sqrt{2}\cosh (\theta +\sigma _{21})}{\cosh 2(\theta +\sigma _{21})}
& \frac{-2\sqrt{2}\cosh \theta }{\cosh 2\theta } & \frac{-2}{\cosh \theta }
& \frac{-2\sqrt{2}\cosh \theta }{\cosh 2\theta } \\ 
\frac{1}{\cosh (\theta +\sigma _{21})} & \frac{-1}{\cosh \theta } & \frac{-2%
\sqrt{2}\cosh \theta }{\cosh 2\theta } & \frac{-1}{\cosh \theta }
\end{array}
\right) \,.  \label{kernels}
\end{equation}
We are content here with the presentation of the analytic approximations for
the plateaux. As we are dealing with functions strongly peaked at the
origin, we can approximate the TBA-equations and obtain the effective
central charge in the deep UV from the constant TBA-equations (\ref{ceff}).
The $N$-matrix involved in there reads 
\begin{equation}
N_{Sp(4)_{2}}=\frac{1}{2\pi }\int\limits_{-\infty }^{\infty }d\theta \text{ }%
\Phi _{Sp(4)_{2}}(\theta )=\left( 
\begin{array}{cccc}
0 & 1/2 & 1 & 1/2 \\ 
1/2 & -1/2 & -1 & -1/2 \\ 
1 & -1 & -1 & -1 \\ 
1/2 & -1/2 & -1 & -1/2
\end{array}
\right) .  \label{nmatrix}
\end{equation}
One can easily check that the same matrix (\ref{nmatrix}) is also obtained
when specializing the general formula given in \cite{CK}. The corresponding
constant TBA solutions of (\ref{ceff}) are 
\begin{equation}
x_{1}^{1}=3,\quad x_{1}^{2}=x_{1}^{4}=2/3,\quad x_{1}^{3}=4/5,
\end{equation}
which yield the Virasoro effective central charge 
\begin{equation}
c_{\text{eff}}=\frac{6}{\pi ^{2}}\left[ \mathcal{L}\left( \frac{3}{4}\right)
+2\mathcal{L}\left( \frac{2}{5}\right) +\mathcal{L}\left( \frac{4}{9}\right) 
\right] \,=2\,,
\end{equation}
in complete agreement with the expectations (\ref{cdel}), for $k=2$, $\ell =2
$, $h=4$ and $h^{\vee }=3$. Let us now apply the decoupling rule to (\ref
{so5}). We expect\footnote{%
Here the arrow in the Dynkin diagram is not related to the sign of $\sigma $
and has the usual meaning, that is, pointing from the long to the short root.%
}

\unitlength=1.0cm 
\begin{picture}(15.20,2.3)(-1.8,-0.8)
\put(0,1.00){\circle*{0.2}}
\put(-0.65,.50){$SU(2)$}
\put(0.00,1.06){\line(1,0){0.9}}
\put(0.00,0.92){\line(1,0){0.9}}
\put(0.25,0.99){\line(2,-1){0.4}}
\put(0.25,0.99){\line(2,1){0.4}}
\put(1.00,1.00){\circle*{0.2}}
\put(0.9,0.50){$SU(4)$}
\put(5.0,0.85){\normalsize $ Sp(4)_2$}
\put(10.5,0.8){$2$} 
\put(-2.0,0.00){$\rightarrow \sigma_{12}$}
\put(0,0.00){\circle*{0.2}}
\put(-0.65,-0.50){$SU(2)$}
\put(1.00,0.00){\circle*{0.2}}
\put(0.9,-0.50){$SU(4)$}
\put(5.0,-0.15){$ SU(2)_2 \otimes SU(2)_4$}
\put(10.5,-0.2){$1.5$}
\end{picture}

\noindent Indeed taking the limit $\sigma _{12}\rightarrow \infty $ in (\ref
{kernels}) gives the $N$-matrix 
\begin{equation}
\lim_{\sigma _{12}\rightarrow \infty }N_{Sp(4)_{2}}=\left( 
\begin{array}{cccc}
0 & 0 & 0 & 0 \\ 
0 & -1/2 & -1 & -1/2 \\ 
0 & -1 & -1 & -1 \\ 
0 & -1/2 & -1 & -1/2
\end{array}
\right) .  \label{nmatrix2}
\end{equation}
We note that particle $(1,1)$ has completely decoupled from the other
particles. The solutions of the constant TBA equations are then 
\begin{equation}
x_{1}^{1}=1,\quad x_{1}^{2}=x_{1}^{4}=1/2,\quad x_{1}^{3}=1/3,
\end{equation}
which gives 
\begin{equation}
c_{\text{eff}}=\frac{6}{\pi ^{2}}\left[ \mathcal{L}\left( \frac{1}{2}\right)
+2\mathcal{L}\left( \frac{1}{3}\right) +\mathcal{L}\left( \frac{1}{4}\right) 
\right] \,=3/2,
\end{equation}
in complete agreement with our expectations. Taking now $m_{1}$ or $m_{2}$
to infinity yields $SU(2)_{4}$ or $SU(2)_{2}$, respectively. This is what is
predicted from the relations (3.9) in \cite{Kuniba}, which makes the
proposal in \cite{CK} natural.

\section{Conclusion}

We have proposed a new bootstrap principle, which involves also unstable
particles in the scattering processes. So far, only primary unstable
particles were analyzed in the literature, which occurred merely as side
products in the scattering processes of stable particles. Our proposal goes
beyond this and has predictive power, as it allows to evaluate the mass
spectrum of further unstable particles, such as secondaries, tertiaries,
etc. In addition, it explains the degeneracy of tertiary unstable particles
and possibly of higher generations for certain theories.

We commented on the general Lie algebraic picture, which underlies the
construction of all known scattering theories which involve unstable
particles in their spectrum. Within this picture we propose various new
scattering matrices for combinations of algebras not explored so far.
Furthermore, we proposed a consistent way to go beyond scaling theories of
statistical models and to incorporate an effective coupling constant. The
structure of the models obtained in this way can be enhanced by a
complexification of the coupling constant and hence combines the ``roaming
models'' with ``Lie algebraic'' ones. A more detailed analysis of these
latter models would be very interesting and lead to yet unknown staircase
patterns in the RG flow.

For all integrable scattering theories which are of the general form as
discussed in section 3,we formulate a new Lie algebraic decoupling rule
which predicts the fixed points of the RG flow from the ultraviolet to the
infrared. The decoupling rule is in agreement with the bootstrap
construction. Still, it would be highly desirable to derive the decoupling
rule more rigorously from first principles.

Our proposals are additionally confirmed by a TBA analysis, which reproduces
the bootstrap prediction of the mass spectrum of unstable particles
including those which are degenerate and hence invisible in the RG flow. For
the HSG models the implications are that each positive root is associated to
a particle. All predictions from our decoupling rule are in perfect
agreement with the outcome of our TBA analysis, even for involved high rank
algebras such as $E_{6}$, for higher level and the non-simply laced
case.\medskip

\noindent \textbf{Acknowledgments: } O.A.~C-A. and A.F. are grateful to the
Deutsche Forschungsgemeinschaft (Sfb288) for financial support. J.D. is
grateful to the Studienstiftung des deutschen Volkes for financial support.

\end{document}